\documentclass[11pt]{article}
\usepackage{fullpage,graphicx,amsmath,amssymb,cancel,float,color,mathrsfs,amsfonts,eufrak,amsthm}
\usepackage{bbm}
\usepackage{subfigure}
\usepackage{cite}
\usepackage{enumerate}
\usepackage{blindtext}
\usepackage{abstract}

\newtheoremstyle{remboldstyle}
  {}{}{}{}{\bfseries}{.}{.5em}{{\thmname{#1 }}{\thmnumber{#2}}{\thmnote{ (#3)}}}
\theoremstyle{remboldstyle}

\newcommand{\drop}[1]{}

\def\tr{\operatorname{tr}}

\def\div{\operatorname{div}}

\begin{document}
\title{A Three-dimensional Study of Coupled Grain Boundary Motion with Junctions}
\author{Anup Basak and Anurag Gupta\thanks{Corresponding author}}
\date{{\small Department of Mechanical Engineering, 
Indian Institute of Technology, Kanpur 208016, 
India\\ \today}}
\maketitle
\begin{abstract}
A novel continuum theory of 
incoherent interfaces with triple junctions is applied to study three-dimensional coupled grain boundary (GB) motion in 
polycrystalline materials. The kinetic relations for grain dynamics, relative sliding and migration of the boundary, 
and junction evolution are developed. In doing so a vectorial form of the geometric coupling factor, which relates the tangential motion at the GB to the migration, is also obtained. Diffusion along the GBs and the junctions is allowed so as to prevent nucleation of voids and overlapping of material 
near the GBs. The coupled dynamics has been studied in detail for two bicrystalline and one tricrystalline arrangements. The first bicrystal consists of two rectangular 
grains separated by a GB, while the second is composed of a spherical grain 
embedded inside a larger grain. The tricrystal has an arbitrary shaped grain embedded inside a much larger
bicrystal made of two rectangular grains. In all these cases, analytical solutions are obtained wherever possible while emphasizing the role of various kinetic coefficients during the coupled motion.

\vspace{4mm}\noindent {\bf Keywords:} Incoherent interfaces; Triple junction; Coupled grain boundary motion; 
 Geometric coupling factor; Nanocrystalline materials
\end{abstract}

\section{Introduction}
\label{introductions_3D}
We develop a thermodynamically consistent continuum framework to study three-dimensional (3D) coupled grain boundary (GB) motion in the presence of triple junctions. A GB is modelled as a sharp incoherent interface connected to other GBs at junction curves. The irreversible dynamics at a GB is governed by its normal motion (GB migration) and a relative tangential sliding of the adjacent grains. The latter can arise due to the inter-granular viscous sliding, possibly as a result of the twist component of the GB, or/and as a result of coupling with 
GB migration \cite{cahn1}. In polycrystalline materials with rigidly deforming grains, as will be assumed presently, the sliding can be decomposed into a relative translation and a relative rotation between the adjacent grains. On the other hand, the irreversible dynamics at a junction is governed by the motion of the non-splitting junction curve. The presence of junctions can significantly influence the overall dynamics of all the GBs and the grains in their neighborhood, for instance by inducing drag or altering diffusive flux \cite{basak2}. The coupled motion, which requires sliding to be necessarily coupled with GB migration,  is the dominant mechanism for both grain coarsening and plastic deformation in nanocrystalline (NC) materials with average grain 
size of the order of few tens of nanometers (hence a large volume fraction of  GBs and triple junctions)\cite{koch1, wang1}. This is unlike coarse-grained materials where GB migration and dislocation dynamics dominate grain coarsening and plastic deformation, respectively. 
The coupled motion has recently been studied theoretically \cite{cahn1,taylor1,basak1,basak2}, experimentally \cite{gorkaya2}, and with molecular simulations \cite{trautt3}. Although some of these studies have included the effect of junction dynamics \cite{trautt3, basak2}, all of them are restricted to two-dimensional grains and hence applicable only to polycrystals where each grain is columnar and identical 
in cross-section along the length direction; such a restriction requires the GB to have only tilt, and no twist, character. 

The main contributions of this paper include:

\noindent (i) A 3D thermodynamic formalism including diffusion to deal with incoherent interfaces with junctions (Section \ref{balance_laws}). Junctions have been previously studied in the context of continuum thermodynamics but only with coherent interfaces and without diffusion \cite{simha2, mariano1}. On the other hand, thermodynamics of incoherent interfaces has been explored earlier without considering junctions \cite{cermelli2}. All of these works were based on the framework of configurational mechanics. Our treatment, while extending to junctions with incoherent interfaces, takes an alternate viewpoint where we do not regard the configurational forces to be fundamentally on the same footing as standard forces (with their own balance laws etc.). We introduce configurational forces in our formalism as mechanisms of internal power generation so as to ensure that the excess entropy production is restricted to interfaces and junctions. A 2D version of this formalism was recently presented by the authors \cite{basak2}.

\noindent(ii) Deriving kinetic relations for coupled GB motion in three dimensions (Section \ref{examples_3D_kinetics}). We extend earlier models of coupled GB motion to a 3D setting. The first kinetic relations for the coupled motion were proposed by Cahn and Taylor \cite{cahn1, taylor1} which were restricted to two-dimensions and only bicrystalline arrangements (hence no junctions). They also ignored the possibility of relative translation of grains while considering sliding at the GB only due to the relative rotation. More recently, the present authors have extended the model to include junctions and relative translation but still restricting themselves to two dimensions \cite{basak2}. 

\noindent(iii) Formulation of a vectorial geometric coupling factor (Section \ref{coupling_factor_3D_deriv}). The coupling between the tangential and the normal motion of the GB is purely geometric and depends on the measure of incoherency at the boundary  \cite{cahn1, cahn2, cahn3}. The incoherency is quantified by the net Burgers vector (given by Frank-Bilby relation) or equivalently by the interfacial dislocation density. The GBs in the present 3D framework generally have a mixed character with both tilt and twist components. The coupling factor for a high angle planar symmetric tilt boundary, derived previously by Cahn et al. \cite{cahn2, cahn3}, therefore needs to be extended to include multiple sets of edge and screw dislocation arrays. The coupling factor now derived is a vectorial quantity rather than a scalar as has been the case in the earlier studies.  

Our derivation for kinetic relations is based on the following
assumptions: (a) the individual grains experience only rigid deformations (i.e. translations and rotations), (b) the shape accommodation required for preventing void-formation and interpenetration of the material in proximity of the GBs, during relative tangential motion between the grains, is accomplished by
diffusion across as well as along the GBs and also along the junction curves; (c) the velocities associated with various GBs, grains, and 
junctions remain much smaller than the speed of sound in the material. The inertial effects are therefore ignored; (d) 
the grains are free of defects and all the lattice imperfections are concentrated at the GBs and junctions (this is reasonable for NC materials with their small grain size); and (e) no additional stress fields are present at the interface and the junction.
The GBs are considered to be orientable surfaces (of arbitrary shapes) with five macroscopic degrees of freedom which include three misorientation angles and two independent variables describing the 
orientation of the GB. The junctions are arbitrary 3D space curves with varying curvature, normal, binormal, torsion 
etc. The excess energy density of a GB is assumed to depend on the five parameters mentioned above, while  the excess energy density of a junction is assumed to depend only on the unit tangent associated with the junction curve.

The paper has been organized in the following manner. In Section \ref{kinematics_evolving_gb} we briefly 
introduce various kinematic and integral relations required for our study.
We derive the essential balance laws and dissipation inequalities in Section \ref{balance_laws}.
A generalized derivation of the vectorial geometric coupling factor 
has been presented in Section \ref{coupling_factor_3D_deriv}. In Section \ref{examples_3D_kinetics} we apply our theory to derive the kinetic relations for GB motion, grain dynamics, and junction motion for two bicrystalline and one tricrystalline arrangements. The phenomenological kinetic equations are motivated from the dissipation inequalities derived from the second law of thermodynamics in confirmation with other
standard balance laws of continuum physics. One bicrystal has two rectangular grains separated 
by a low angle planar mixed GB, while the other has a spherical grain embedded inside a larger grain. 
The tricrystal constitutes of an arbitrary 3D grain which is completely embedded inside a bicrystal 
consisting of two large rectangular grains. We conclude our study with a discussion on some open directions in 
Section \ref{conclusion_junctions_3D}.

\section{Kinematics}
\label{kinematics_evolving_gb}
\begin{figure}[t!]
\centering 
\subfigure[]{
  \includegraphics[width=2.8in, height=1.4in] {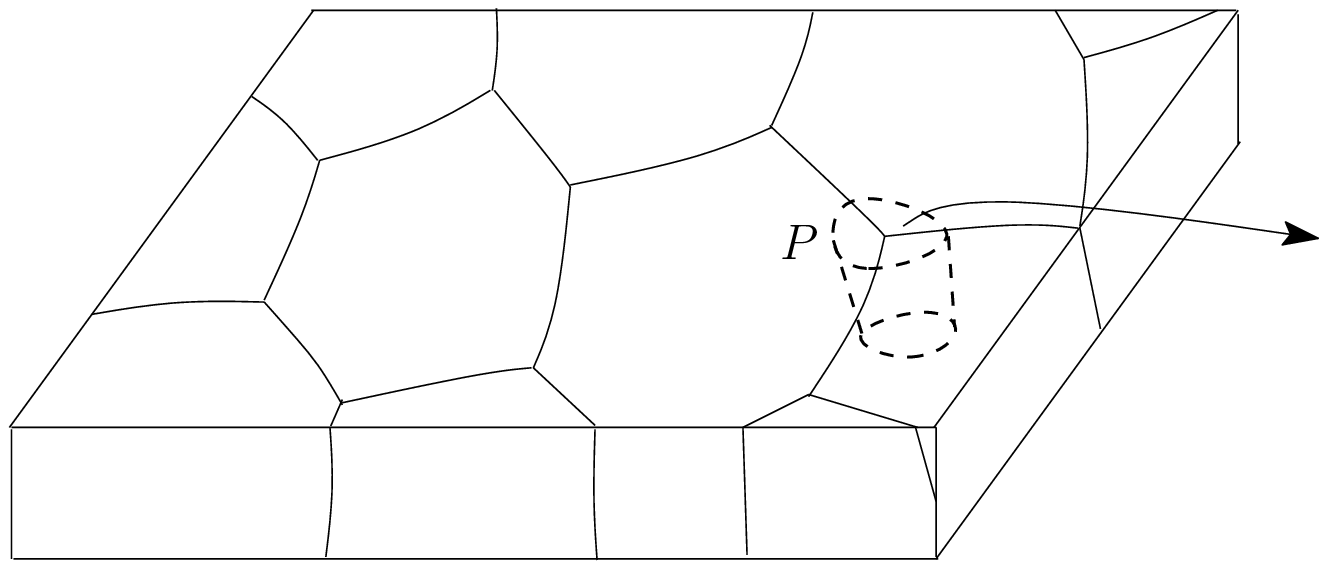}
	\label{polycrystal_3D}}
\hspace{10mm}
    \subfigure[]{
    \includegraphics[width=2.7in, height=2.2in] {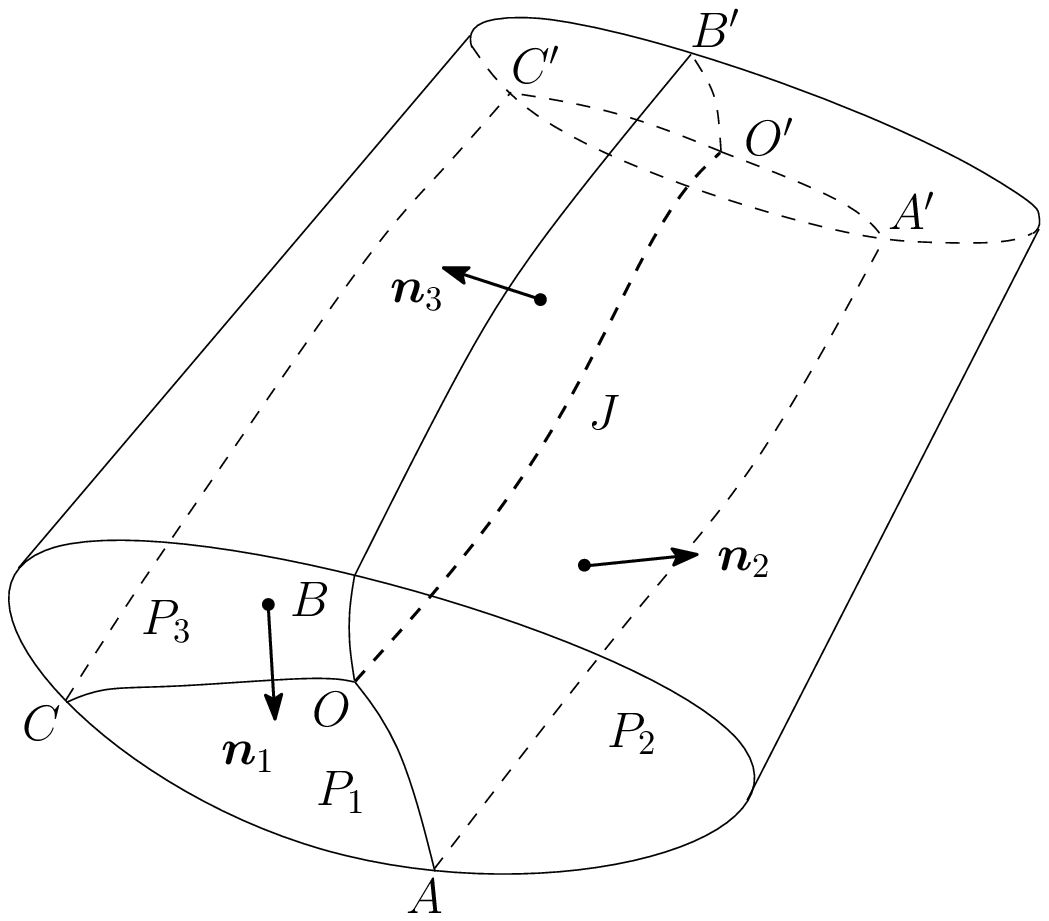}
    \label{control_volume_junc}}
 \caption{(a) Schematic of a polycrystal in 3D. (b) The region $P$ containing three subregions 
$P_1$, $P_2$, and $P_3$. The GBs $OO'A'A$, $OO'B'B$, 
and $OO'C'C$ (with normals 
${\boldsymbol n}_1$, ${\boldsymbol n}_2$, and ${\boldsymbol n}_3$, as shown), are denoted by $\Gamma_1$, $\Gamma_2$, and $\Gamma_3$, respectively. The curve $OO'$ is
the triple junction $J$.}
\end{figure}

We consider a region $P$,  as shown in Figure \ref{control_volume_junc}, taken 
out from a polycrystalline arrangement depicted in Figure \ref{polycrystal_3D}. It contains three subgrains 
$P_1$, $P_2$, and $P_3$, three smooth GBs $\Gamma_1$, $\Gamma_2$, and $\Gamma_3$, and a smooth 
junction curve $J$. The normal ${\boldsymbol n}_i$ to $\Gamma_i$ is chosen such that it points inside $P_i$,
where $i=1,2,3$. We denote the position vector of a point by ${\boldsymbol x}$ and the time by $t$. 
The grains are oriented differently with respect to a fixed coordinate. 
 
Let ${\boldsymbol A}$ and ${\boldsymbol B}$ be two second order tensors. The derivative of a scalar valued differentiable function of tensors, say  $G({\boldsymbol A})$,  is a tensor $\partial_{\boldsymbol A} G$ defined by
\begin{equation}
G({\boldsymbol A} + {\boldsymbol B}) = G({\boldsymbol A}) + \partial_{\boldsymbol A} G \cdot {\boldsymbol B} + o(|{\boldsymbol B}|), 
\end{equation}
where $o(|{\boldsymbol B}|)/
|{\boldsymbol B}| \rightarrow 0$ as $|{\boldsymbol B}| \rightarrow 0$; the norm of a tensor is defined as $|{\boldsymbol B}|^2 = {\boldsymbol B}\cdot{\boldsymbol B}$. Similar definitions can be made for vector and tensor valued
differentiable functions (of scalars, vectors, and tensors). The derivative of a field defined over $P$ with respect to the position vector is denoted by the gradient operator $\nabla$.  

\subsection{Bulk fields}
Let $f$ be a piecewise smooth bulk field which is discontinuous across  $\Gamma_i$ and singular at $J$.
We denote the jump of $f$ across $\Gamma_i$ as $[\![f]\!]=f^+-f^-$, where $f^+$ is the limiting value 
 of $f$ at ${\boldsymbol x}\in\Gamma_i$ from  the grain into which ${\boldsymbol n}_i$ 
 points and $f^-$ is the limiting value from the other grain. If $f_1$ and 
 $f_2$ are two piecewise continuous functions across $\Gamma_i$, then 
 $[\![f_1f_2]\!] = [\![f_1]\!]\langle f_2\rangle+\langle f_1\rangle[\![f_2]\!]$, 
 where $\langle f \rangle = (f^+ +f^-)/2$ is the average of $f^+$ and $f^-$.
To deal with the singularity of the field at the junction we carry out 
our analysis in a punctured region $P_\epsilon$ obtained by excluding a small tube 
$T_\epsilon$ of radius $\epsilon$ from $P$ in the neighborhood of the junction, cf. \cite{simha2}. The outward normal to the boundary of the tube
$\partial T_\epsilon$ is denoted by ${\boldsymbol m}$. The boundary 
$\partial{T}_\epsilon$ moves with a velocity ${\boldsymbol u}$.
 
 We assume that $f$ satisfies the limit
\begin{equation}
\int_{P} f dv = \lim_{\epsilon\to 0}\int_{{P}_\epsilon}f dv,
 \label{regularity_field_junc}
\end{equation}
where $dv$ is an infinitesimal volume element of $P$. Using the standard transport relations for the bulk 
quantities it can be shown that \cite{simha2}
\begin{equation}
\frac{d}{dt}\int_{P}f dv = \int_{P}(\dot{f}+f\div {\boldsymbol v}) dv-\sum_{i=1}^3
 \int_{\Gamma_i}[\![f U_i]\!] da-\lim_{\epsilon\to 0}\int_{\partial{T}_\epsilon
 }f ~({\boldsymbol u}-{\boldsymbol v})\cdot{\boldsymbol m} da,
\label{bulk_transport2}
\end{equation}
where the superposed dot denotes the material time derivative, $\div$ is the divergence operator, 
${\boldsymbol v}$ is the particle velocity, $V_i$ is the interfacial normal velocity, 
$U_i=V_i-{\boldsymbol v}\cdot{\boldsymbol n}_i$ is the relative normal velocity of the interface, 
and $da$ is an infinitesimal area element of a surface. Let ${\boldsymbol a}$ and ${\boldsymbol A}$
denote piecewise-smooth vector and tensor fields, defined in $P$, which are singular at the junction.
The divergence theorem requires \cite{simha2}
\begin{equation}
\int_{P}\div {\boldsymbol a} dv = \int_{\partial{P}}{\boldsymbol a}\cdot{\boldsymbol m} da
-\sum_{i=1}^3\int_{\Gamma_i}[\![{\boldsymbol a}]\!] \cdot{\boldsymbol n}_i da
-\lim_{\epsilon\to 0}\int_{\partial{T}_\epsilon}{\boldsymbol a}\cdot{\boldsymbol m} da~\text{and}
 \label{bulk_divergence2}
\end{equation}
 \begin{equation}
\int_{P}\div{\boldsymbol A} dv =
\int_{\partial{P}}{\boldsymbol A}{\boldsymbol m} da
-\sum_{i=1}^3\int_{\Gamma_i}[\![{\boldsymbol A}]\!]{\boldsymbol n}_i da
-\lim_{\epsilon\to 0}\int_{\partial{ T}_\epsilon}{\boldsymbol A}{\boldsymbol m} da.
 \label{bulk_divergence3}
\end{equation}

\subsection{Interfacial fields}
Consider an orientable surface $\Gamma$ (subscript $i$ is presently dropped), with boundary $\partial \Gamma$,  and let ${\boldsymbol n}$ and $V$ be the associated unit normal field and normal velocity field, respectively.
The surface gradient of a scalar field $g$, vector field $ {\boldsymbol g}$, and tensor field $ {\boldsymbol G}$, all smoothly defined over $\Gamma$, are defined as 
\begin{equation}
\nabla^S g={\boldsymbol P}(\nabla g),~{\nabla}^S {\boldsymbol g}
=(\nabla {\boldsymbol g}){\boldsymbol P},~\text{and}~
\nabla^S {\boldsymbol G}=(\nabla {\boldsymbol G}) {\boldsymbol P},
\label{surf_grad1}
\end{equation}
respectively, 
where $ {\boldsymbol P}={\boldsymbol I}-{\boldsymbol n}\otimes{\boldsymbol n}$ 
is the projection tensor (${\boldsymbol I}$ is the 3D identity tensor and $\otimes$ denotes the dyadic product); while calculating $\nabla g$ (etc.) one has to use a smooth extension of $g$ in a small neighborhood of $\Gamma$. The surface divergence of these fields are defined by
\begin{equation}
\div^S{\boldsymbol g}= \tr (\nabla^S  {\boldsymbol g}) ~\text{and}
 ~ {\boldsymbol k}\cdot \div^S {\boldsymbol G}=\div^S( {\boldsymbol G}^T
{\boldsymbol k}),
 \label{surf_div_vect}
\end{equation}
for all constant vectors ${\boldsymbol k}$, where $\tr$ 
represents the trace operator and the superscript $T$ stands for the transpose. The surface 
Laplacian of $g$ is given by
\begin{equation} ~
\Delta^S g = \div^S(\nabla^S g).
 \label{surface_laplacian}
\end{equation}
The curvature tensor field $ {\boldsymbol L}$ and the total curvature $\kappa$ 
associated with $\Gamma$ are defined as 
\begin{equation}
 {\boldsymbol L}= -\nabla^S {\boldsymbol n}~\text{and} ~ \kappa= \tr {\boldsymbol L},
 \label{curvature_tensor} 
\end{equation}
respectively.

 Let ${\boldsymbol t}$ be the outward unit normal to the closed curve $\partial \Gamma$ such that ${\boldsymbol n}\cdot {\boldsymbol t}=0$. When $ {\boldsymbol G}$ and $ {\boldsymbol g}$ satisfy 
$ {\boldsymbol G}{\boldsymbol n}={\boldsymbol 0}$ and $ {\boldsymbol g}\cdot{\boldsymbol n}= 0$, respectively,
the surface divergence theorem yields \cite{gupta1}
\begin{equation}
\int_{\partial{\Gamma}} {\boldsymbol G} {\boldsymbol t} dl
=\int_{{\Gamma}}\div^S {\boldsymbol G}  da
 ~\text{and} ~ \int_{{\partial\Gamma}} {\boldsymbol g}\cdot {\boldsymbol t} dl
=\int_{\Gamma}\div^S {\boldsymbol g}  da,
 \label{divergence_thms}
\end{equation}
where $dl$ is an infinitesimal line element along a curve. 

The normal time derivative of $g$ following $\Gamma$ is given by \cite{gupta1} 
\begin{equation}
\overset{\circ}{g} = \dot{g}+ V \nabla g\cdot{\boldsymbol n},
 \label{normal_time_derivative3}
 \end{equation}
which is the rate of change of $g$ as experienced by an observer sitting on the moving surface $\Gamma$. The first term 
indicates the local rate of change of $g$ at a fixed material position, while the second term represents the 
rate of change of $g$ due to influx of particles along ${\boldsymbol n}$ as the interface moves with velocity $V$. The following identities can be readily verified \cite{gurtin2}:
\begin{equation}
\overset{\circ}{\boldsymbol n}=-\nabla^S V~\text{and}~\overset{\circ}{\boldsymbol L}= - \nabla^S \overset{\circ}{\boldsymbol n} -{\boldsymbol L} \overset{\circ}{\boldsymbol n} \otimes {\boldsymbol n} + V {\boldsymbol L}^2. \label{normal_time_derivativenl}
\end{equation}

On the other hand, the intrinsic 
time derivative of $g$ following $\partial\Gamma$ is given by \cite{gurtin2}
\begin{equation}
\overset{\square}{g} = \dot{g}+\nabla g \cdot {\boldsymbol w},
\label{time_derivative_boundary2}
\end{equation}
where ${\boldsymbol w}$ is the intrinsic velocity of $\partial\Gamma$, such that
\begin{equation}
{\boldsymbol w}= V{\boldsymbol n}+W{\boldsymbol t},
 \label{boundary_velocity_gammai}
\end{equation}
and $W$ is the velocity of $\partial \Gamma$ along ${\boldsymbol t}$. According to \eqref{time_derivative_boundary2}
the rate of change of $g$ following $\partial\Gamma$ is equal to the sum of the rate of 
change of $g$ following $\Gamma$ and a term representing 
the change in $g$ due to the incoming particles from the neighborhood along the tangential direction 
${\boldsymbol t}$. 

We will need the following transport theorem for $\Gamma$ such that a part of $\partial \Gamma$ intersects with $\partial P$ and the rest with $J$ \cite{simha2}:
\begin{equation}
\frac{d}{dt}\int_{\Gamma}g da 
= \int_{\Gamma}(\overset{\circ}{g}-g\kappa V) da
+\int_{\Gamma\cap\partial{P}}g W dl+\int_{J}g {\boldsymbol q}_p\cdot{\boldsymbol t} dl,
\label{gb_transport2}
\end{equation}
where ${\boldsymbol q}_p$ is the intrinsic (independent of the parametrization) velocity of the junction.

\subsection{Junction fields}
Let $\delta$ and $\iota$ denote the the terminal 
points of the junction curve $J$, and let ${\boldsymbol l}$ be the unit tangent to the curve such that it is directed towards $\iota$. The normal and binormal vectors associated with $J$ 
are denoted by ${\boldsymbol\nu}$ and ${\boldsymbol b}$, respectively. The projection tensor 
${\boldsymbol Q} = {\boldsymbol I}-{\boldsymbol\nu}\otimes{\boldsymbol\nu}-{\boldsymbol b}
\otimes{\boldsymbol b}= {\boldsymbol l}\otimes{\boldsymbol l}$ maps any vector 
on to the tangential direction of $J$. The intrinsic velocity field of the junction, which can be decomposed as
\begin{equation}
{\boldsymbol q}_p=q_\nu{\boldsymbol\nu}+q_b{\boldsymbol b},
\label{velocity_junction_proj}
\end{equation}
is such that  $({\boldsymbol I}
-{\boldsymbol Q}){\boldsymbol u}\to{\boldsymbol q}_p ~{\rm as}~  \epsilon\to 0$  \cite{simha2}.
The velocity of terminal points is denoted as $\hat{\boldsymbol q}$ such that 
\begin{equation}
\hat{\boldsymbol q} = {\boldsymbol q}_p + \hat{q}_l {\boldsymbol l}
\label{velocity_junction_endpts}
\end{equation} 
at the respective end points.  
The intrinsic time derivative of a scalar field defined on the junction curve, say $\chi$  is given by  (compare with \eqref{time_derivative_boundary2})
\begin{equation}
 \overset{\star}{\chi} =\dot{\chi} + \nabla \chi \cdot  {\boldsymbol q}_p.
\label{normal_derivative_junction}
\end{equation}
The transport theorem associated with $\chi$ is given by \cite{simha2}
\begin{equation}
\frac{d}{dt}\int_{J}\chi dl =\int_{J}( \overset{\star}{\chi}-\chi\kappa_Jq_\nu) dl
+(\chi \hat{q}_l)_{\delta}^{\iota},
\label{junction_transport1}
\end{equation}
where $\kappa_J$ is the curvature of the junction curve. The gradient of $\chi$ along the junction curve is defined as $\nabla^J \chi = {\boldsymbol Q} \nabla \chi$. Similarly for a vector field defined on $J$ we introduce $\nabla^J {\boldsymbol q}_p =  (\nabla {\boldsymbol q}_p){\boldsymbol Q}$. We note the identity \cite{simha2}
\begin{equation}
 \overset{\star}{\boldsymbol l} = (\nabla^J {\boldsymbol q}_p){\boldsymbol l} + q_\nu \kappa_J {\boldsymbol l}.
\label{normal_derivative_tan}
\end{equation}

It is useful to decompose an integral over the tube surface around the junction as
\cite{simha2}
 \begin{equation}
\lim_{\epsilon\to 0}\int_{\partial{T}_\epsilon}{\boldsymbol a} da = 
\int_{J}\left[\lim_{\epsilon\to 0}\int_{{C}_\epsilon} {\boldsymbol a}  dl\right]dl,
 \label{integral_appox_junc}
\end{equation}
where $\partial{ T}_\epsilon$ is the envelope of the circles ${ C}_\epsilon$ of radius $\epsilon$.

\section{Balance laws and dissipation}
\label{balance_laws}
 
We now obtain the consequences of balance of mass and momentum, as well as obtain local dissipation inequalities in the bulk, at the interface, and at the junction, while restricting to the assumptions enlisted in Section  \ref{introductions_3D}.

\subsection{Balance of mass}
\label{balancemass_junc}
The rate of change of total mass in $P$ is balanced by the mass 
transport into the region
 via bulk diffusion across $\partial{P}$, GB diffusion at the edge $\Gamma_i\cap\partial{P}$, and 
diffusion at $\iota$ and $\delta$. Neglecting excess mass densities of the GBs and the junction, the mass balance can be 
written as
\begin{equation}
\frac{d}{{d}t}\int_{P} \rho dv = -\int_{\partial{P}}{\boldsymbol j}\cdot{\boldsymbol m} da
-\sum_{i=1}^3\int_{\Gamma_i\cap\partial{P}}{\boldsymbol h}_i\cdot {\boldsymbol t}_i dl
-(h_J)_{\delta}^{\iota},
\label{bomint_junc}
\end{equation}
where $\rho$ is the mass density of the bulk grain, ${\boldsymbol j}$ is the bulk diffusional flux, 
${\boldsymbol h}_i$ is the tangential diffusional flux on $\Gamma_i$, and $h_J$ is the diffusional flux 
along the junction. 
Using transport theorem \eqref{bulk_transport2} and divergence theorems \eqref{bulk_divergence2} and \eqref{divergence_thms}$_2$, and localizing the result owing to the arbitrariness of $P$,  we can obtain the following local equations:
\begin{equation}
\dot\rho + \rho \div {\boldsymbol v}+\div{\boldsymbol j}=0 ~
\forall{\boldsymbol x}\in{P}_i,
\label{bom_junc}
\end{equation}
\begin{equation}
[\![\rho U_i ]\!]=[\![{\boldsymbol j}]\!]\cdot{\boldsymbol n}_i+\div^S{\boldsymbol h}_i ~\forall{\boldsymbol x}\in{\Gamma}_i,~\text{and}
\label{boms_junc}
\end{equation}
\begin{equation}
\lim_{\epsilon\to 0} \int_{{ C}_\epsilon} \rho  ({\boldsymbol u}-{\boldsymbol v}) \cdot{\boldsymbol m} dl = \nabla^J h_J\cdot {\boldsymbol l} +\lim_{\epsilon\to 0} \int_{{ C}_\epsilon} {\boldsymbol j}\cdot{\boldsymbol m} dl- \sum_{i=1}^3\left({\boldsymbol h}_i\cdot{\boldsymbol t}_i\right)_J
 ~\forall{\boldsymbol x}\in J,
\label{bomjun_junc}
\end{equation}
where we have used the 
 limit $\lim_{\epsilon\to 0}\sum_{i=1}^3\int_{\partial{T}_\epsilon\cap\Gamma_i}
{\boldsymbol h}_i\cdot {\boldsymbol t}_i dl=\sum_{i=1}^3\int_J
\left({\boldsymbol h}_i\cdot{\boldsymbol t}_i\right)_J dl$.

\subsection{Balance of linear momentum}
\label{balancelmom}
Neglecting inertia and body forces, and assuming absence of interfacial and junction stress fields, the balance of linear momentum is given by 
\begin{equation}
\int_{\partial  P} {\boldsymbol \sigma} {\boldsymbol m} da={\boldsymbol 0}, 
\label{blmint_junc}
\end{equation}
where ${\boldsymbol\sigma}$ is the symmetric Cauchy stress tensor.
Using \eqref{bulk_divergence3}  the following local equations are readily obtained \cite{simha2}:
\begin{equation}
\div {\boldsymbol \sigma} = {\boldsymbol 0}~\forall{\boldsymbol x}\in{P}_i,
\label{blm_junc}    
\end{equation}
\begin{equation}
[\![{\boldsymbol\sigma}]\!]{\boldsymbol n}_i = {\boldsymbol 0}
~\forall {\boldsymbol x} \in {\Gamma}_i,~\text{and}
\label{blms_junc}  
\end{equation}
\begin{equation}
\lim_{\epsilon\to 0}\int_{C_\epsilon}{\boldsymbol \sigma}{\boldsymbol m} dl={\boldsymbol 0}
~\forall {\boldsymbol x} \in J.
\label{blmjun_junc}
\end{equation}
According to \eqref{blms_junc} the traction field is continuous across the GBs, whereas \eqref{blmjun_junc} requires that the net force acting at each circular region $C_\epsilon$ is zero in
the limit $\epsilon\to 0$, although the stress field
can still be singular at the junction (weak singularity).

\subsection{Dissipation inequality}
\label{dissipation_ineq_junc}
Let $\Psi$ be the bulk free energy density, $\gamma_i$ the interfacial free energy per unit area of  $\Gamma_i$, and $\eta$  the free energy per unit length of $J$. For an isothermal environment, the mechanical version of the second law requires the rate of change of the total free energy $P$ to be less than or equal to the total power input into $P$ \cite{gupta1, fried1}, i.e.
 \begin{eqnarray}
\frac{d}{dt}\left(\int_{P} \Psi dv+\sum_{i=1}^3\int_{\Gamma_i}\gamma_i da+\int_{J}\eta dl\right)
\leq\int_{\partial P} {\boldsymbol \sigma} {\boldsymbol m}\cdot{\boldsymbol v}  da
-\int_{\partial P}\mu {\boldsymbol j}\cdot{\boldsymbol m} da-\sum_{i=1}^3
\int_{\partial{P}\cap\Gamma_i}\mu {\boldsymbol h}_i\cdot{\boldsymbol t}_i dl
\nonumber\\ -(\mu h_J)_{\delta}^{\iota} +\sum_{i=1}^3\int_{\partial P\cap\Gamma_i}({\boldsymbol c}_i
\cdot{\boldsymbol w}_i+{\boldsymbol \tau}_i\cdot\overset{\square}{\boldsymbol n}_i) dl
+\left({\boldsymbol \omega}\cdot\hat{\boldsymbol q}\right)_{\delta}^{\iota},
\label{dissi_ineq_junc1}
\end{eqnarray}
where $\mu$ is the chemical potential. We assume that the chemical potential is continuous across the interface and at the junction (i.e. local chemical equilibrium \cite{fried1}). The first integral on the
right hand side of \eqref{dissi_ineq_junc1} is the power input through the
tractions acting on $\partial{P}$; the next three terms are contribution to power input due to mass flux at
$\partial P$, $\partial P \cap \Gamma_i$, and the end points of $J$, respectively. The last two terms are non-standard; we will discuss their significance before deriving the consequences of \eqref{dissi_ineq_junc1}. 
These terms are required to ensure that there is no excess entropy production 
at the edges $\partial P\cap \Gamma_i$ and at the terminal points of $J$. The excess 
entropy generation is necessarily restricted to the 
GBs and the junction. These additional power input terms are to be considered in \eqref{dissi_ineq_junc1} only
when the edges and the terminal points lie on the surface of an interior part of a body. The precise form of ${\boldsymbol c}_i$, 
${\boldsymbol \tau}_i$, and ${\boldsymbol \omega}$ will depend on the 
constitutive prescriptions for free energies and stress. At this point these are to be understood as agents of power input, in conjugation with the respective intrinsic velocities, so as to ensure that the net entropy generation meets the above mentioned requirement.  Such terms also appear in the framework of configurational mechanics \cite{cermelli2, gurtin2 ,simha2}, where the existence of ${\boldsymbol c}_i$, 
${\boldsymbol \tau}_i$, and ${\boldsymbol \omega}$ is assumed \textit{a priori} as fundamental forces which satisfy certain balance relations. Our treatment (see also \cite{gupta1, basak1, basak2}) is motivated purely from the viewpoint of quantifying excess entropy generation. For this we do not have to consider any additional balance laws other than those which are standard in continuum physics. 

Using transport theorems \eqref{bulk_transport2}, \eqref{gb_transport2}, and \eqref{junction_transport1}, divergence theorems  \eqref{bulk_divergence2} and \eqref{divergence_thms}, and the decomposition \eqref{integral_appox_junc}, we can 
rewrite the inequality  \eqref{dissi_ineq_junc1} as
\begin{equation}
\sum_{a=1}^6 I_a\leq 0,
\label{dissi_ineq_junc2}
\end{equation}
where
\begin{equation}
I_1=\int_{P}\left(\dot\Omega+\Omega\div{\boldsymbol v}+\rho\dot\mu
-{\boldsymbol\sigma}\cdot\nabla{\boldsymbol v}+{\boldsymbol j}\cdot\nabla\mu\right) dv,
\label{dissi_ineq_junc_I1_1}
\end{equation}
\begin{equation}
I_2=\sum_{i=1}^3\int_{P\cap\Gamma_i}\left(\overset{\circ}{\gamma}_i -\gamma_i\kappa_i V_i -[\![U_i{\boldsymbol E}]\!]{\boldsymbol n}_i\cdot{\boldsymbol n}_i
-\langle{\boldsymbol\sigma}{\boldsymbol n}_i\rangle\cdot  {\boldsymbol P}_i[\![{\boldsymbol v}]\!]
+{\boldsymbol h}_i\cdot\nabla^S\mu \right) da,
\label{dissi_ineq_junc_I2}
\end{equation}
\begin{equation}
I_3=\sum_{i=1}^3\int_{\partial{P}\cap\Gamma_i}\left(\gamma_iW_i-{\boldsymbol c}_i\cdot{\boldsymbol w}_i
-{\boldsymbol\tau}_i\cdot\overset{\square}{\boldsymbol n}_i\right) dl,
\label{dissi_ineq_junc_I3}  
\end{equation}
\begin{equation}
I_4=-\int_J\left(\lim_{\epsilon\to 0}\int_{{ C}_\epsilon}\left({\boldsymbol\sigma}{\boldsymbol v}
+\Psi  ({\boldsymbol u}-{\boldsymbol v}) - \mu{\boldsymbol j}\right)\cdot{\boldsymbol m} dl\right)dl,
\label{dissi_ineq_junc_I4}
\end{equation}
\begin{equation}
I_5=\int_J\left(\overset{\star}{\eta}-\eta\kappa_Jq_\nu+ \nabla^J (\mu h_J)\cdot {\boldsymbol l}
+\sum_{i=1}^3(\mu{\boldsymbol h}_i\cdot{\boldsymbol t}_i
+\gamma_i{\boldsymbol q}\cdot{\boldsymbol t}_i)\right)dl,~\text{and}
\label{dissi_ineq_junc_I5}
\end{equation}
\begin{equation}
I_6=\left(\eta\hat{q}_l-{\boldsymbol \omega}\cdot\hat{\boldsymbol q}\right)_{\delta}^{\iota}.
\label{dissi_ineq_junc_I6}
\end{equation}
In obtaining \eqref{dissi_ineq_junc_I1_1} we have used \eqref{bom_junc} and \eqref{blm_junc}, and introduced $\Omega = \Psi-\rho\mu$ (the grand canonical potential). To derive \eqref{dissi_ineq_junc_I2}, on the other hand, we have used \eqref{boms_junc} and \eqref{blms_junc}; here ${\boldsymbol E}=\Omega{\boldsymbol I}-{\boldsymbol\sigma}$ is the bulk Eshelby tensor.

To determine the precise form of ${\boldsymbol c}_i$, ${\boldsymbol\tau}_i$, and ${\boldsymbol\omega}$, and also to obtain the local dissipation inequalities associated with the grains, GBs, and junction, we will now prescribe the constitutive nature of the GB energy and the junction energy. Towards this end, we assume the GB energy to depend on the misorientation between the grains, the normal
 to the GB, and curvature. The former two dependencies are standard in material science literature (cf. Chapter 12 in \cite{read2}). The curvature dependence is primarily introduced to regularize the governing partial differential equations for  capillary driven GB motion, which otherwise become backward
 parabolic and hence unstable in certain ranges (GB spinodals) of the orientations. We follow Gurtin and Jabbour \cite{gurtin2} in assuming the following quadratic dependence of GB energy on curvature:
 \begin{equation}
 \gamma = \hat\gamma({\boldsymbol \Theta}, {\boldsymbol n}) + \frac{1}{2} \epsilon_1 |{\boldsymbol L}|^2 + \frac{1}{2} \epsilon_2 \kappa^2,
 \label{gb_energy_3d_general}
 \end{equation}
where ${\boldsymbol\Theta}$ is the misorientation tensor given by $({\boldsymbol R}^+)^T
{\boldsymbol R}^-$; $\epsilon_1$ and $\epsilon_2$ are scalar constants such that $\epsilon_1 > 0$ and $\epsilon_2 + \epsilon_1/2 >0$. The rotation
${\boldsymbol R}^+$ is the orientation tensor of the grain into which ${\boldsymbol n}$ points, and 
${\boldsymbol R}^-$ is the orientation tensor of the other grain. We introduce
\begin{equation}
{\boldsymbol M} \doteq \partial_{\boldsymbol L} \gamma =  \epsilon_1 {\boldsymbol L} + \epsilon_2 \kappa {\boldsymbol P}.
 \label{gb_energy_L}
\end{equation}
It is easy to see that ${\boldsymbol M}$ is symmetric and satisfies $\boldsymbol{MP} = {\boldsymbol M}$. The junction energy density, on the other hand, is assumed to be a function of the unit tangent along the junction curve \cite{simha2}:
 \begin{equation}
 \eta=\hat\eta({\boldsymbol l}).
 \label{tau_boundary}
 \end{equation}

Substituting \eqref{gb_energy_3d_general} and \eqref{tau_boundary} into \eqref{dissi_ineq_junc2}, and performing a cumbersome but straightforward calculation, yields
\begin{eqnarray}
&&\int_P {\mathfrak D}_b dv
 +\sum_{i=1}^3\int_{\Gamma_i} {\mathfrak D}_{\Gamma_i} da+\int_J {\mathfrak D}_J dl -\sum_{i=1}^3\int_{\partial{ P}\cap\Gamma_i}\Big\{ W_i \left(\gamma_i
-{\boldsymbol M}_i {\boldsymbol L}_i{\boldsymbol t}_i\cdot{\boldsymbol t}_i
-{\boldsymbol c}_i\cdot{\boldsymbol t}_i \right)\nonumber \\
&&
-V_i \left( \left(\div^S
{\boldsymbol M}_i+\partial_{{\boldsymbol n}_i}\hat\gamma_i\right)\cdot{\boldsymbol t}_i+{\boldsymbol c}_i\cdot{\boldsymbol n}_i \right)-\overset{\square}{\boldsymbol n}_i\cdot({\boldsymbol\tau}_i
+{\boldsymbol M}_i {\boldsymbol t}_i)\Big\}dl
\nonumber\\
&&
-\left(\hat{q}_l(\eta-\omega_l)
+\left(({\boldsymbol I}-{\boldsymbol Q})\partial_{\boldsymbol l}\hat\eta
-{\boldsymbol\omega}_p\right)\cdot{\boldsymbol q}_p\right)_{\delta}^{\iota}\geq 0,
\label{total_entropy_generation}
\end{eqnarray}
 where ${\mathfrak D}_b$, ${\mathfrak D}_{\Gamma_i}$, and ${\mathfrak D}_J$ are the
 the entropy generation rates per unit volume of the bulk,  per unit area of $\Gamma_i$, and per unit length
 of $J$, respectively. The expressions for these rates are given in Equations \eqref{local_inequality_bulk}-\eqref{local_inequality_junc1} below. In deriving the above inequality we have also assumed the junctions to be non-splitting, i.e. $V_i={\boldsymbol q}_p\cdot{\boldsymbol n}_i$. 
The first term in the above inequality is the net entropy generation within the grains. The next two terms are excess entropy generation at the GBs and at the junction, respectively. The rest of terms in \eqref{total_entropy_generation} are the entropy production rate at the edges $\Gamma_i\cap\partial P$ and the terminal
 points $\delta$ and $\iota$. We however require that the excess entropy production 
 must not have any contribution from the edges of the GBs and the terminal points of $J$, all of which are a part of $\partial P$. This is reasonable since the entropy generation in $P$ should only be within the grains, at the interfaces, and at the junction. Any additional source should vanish. Consequently
\begin{equation}
{\boldsymbol c}_i\cdot{\boldsymbol t}_i = \gamma_i-{\boldsymbol M}_i
 {\boldsymbol L}_i{\boldsymbol t}_i\cdot{\boldsymbol t}_i,
\label{c_dot_t_3D}
\end{equation}
\begin{equation}
{\boldsymbol c}_i\cdot{\boldsymbol n}_i = - \left(\div^S
{\boldsymbol M}_i+\partial_{{\boldsymbol n}_i}\hat\gamma_i\right)\cdot{\boldsymbol t}_i,
\label{c_dot_n_3D}
\end{equation}
\begin{equation}
{\boldsymbol \tau}_i = -{\boldsymbol M}_i{\boldsymbol t}_i,~\text{and}
\label{c_dot_tau_3D}
\end{equation}
\begin{equation}
{\boldsymbol\omega}=\eta {\boldsymbol l}+({\boldsymbol I}-{\boldsymbol Q})\partial_{\boldsymbol l}\hat\eta.
\label{omega_junc_3D}
\end{equation}

Substituting \eqref{c_dot_t_3D}-\eqref{omega_junc_3D} back into \eqref{total_entropy_generation}, and localizing the result, we obtain the
following local dissipation inequalities:
\begin{equation}
{\mathfrak D}_b = {\boldsymbol\sigma}\cdot\nabla{\boldsymbol v}-(\dot\Omega+\Omega\div{\boldsymbol v})
-\rho\dot\mu-{\boldsymbol j}\cdot\nabla\mu\geq 0 ~\forall{\boldsymbol x}\in P_i,
\label{local_inequality_bulk}
\end{equation}  
\begin{equation}
{\mathfrak D}_{\Gamma_i} = [\![U_i{\boldsymbol E}]\!]{\boldsymbol n}_i\cdot{\boldsymbol n}_i
+\langle{\boldsymbol\sigma}{\boldsymbol n}_i\rangle\cdot {\boldsymbol P}_i[\![{\boldsymbol v}]\!]
-{\boldsymbol h}_i\cdot\nabla^S\mu+f_iV_i-(\partial_{{\boldsymbol\Theta}_i}\hat\gamma_i)
\cdot\dot{\boldsymbol\Theta}_i\geq 0 ~\forall{\boldsymbol x}\in\Gamma_i,~\text{and}
\label{local_inequality_gb1}
\end{equation}
\begin{equation}
{\mathfrak D}_J = {\boldsymbol{\mathcal F}}_J\cdot{\boldsymbol q}_p-\lim_{\epsilon\to 0}
\int_{C_\epsilon}{\boldsymbol E}{\boldsymbol m}\cdot{\boldsymbol v} dl
-h_J (\nabla^J \mu)\cdot{\boldsymbol l}-\sum_{i=1}^3{\boldsymbol\tau}_i
\cdot\overset{\star}{\boldsymbol n}_i\geq 0  ~\forall{\boldsymbol x}\in J,
\label{local_inequality_junc1}
\end{equation}
where 
 \begin{equation}
 f_i=\gamma_i \kappa_i-\div^S(\partial_{{\boldsymbol n}_i}\hat\gamma_i)-{\boldsymbol M}_i\cdot {\boldsymbol L}_i^2- \div^S (\div^S {\boldsymbol M}_i),
 \label{force_normal_velo_3D}
 \end{equation}
\begin{equation}
{\boldsymbol{\mathcal F}}_J=({\boldsymbol I}-{\boldsymbol Q})\left(\lim_{\epsilon\to 0}
\int_{C_\epsilon}{\boldsymbol E}{\boldsymbol m} dl-\sum_{i=1}^3{\boldsymbol c}_i-{\boldsymbol f}_J
\right),~\text{and}
\label{junction_force_expression_3D}
\end{equation}
\begin{equation}
{\boldsymbol f}_J=-\eta\kappa_J{\boldsymbol\nu}+ \nabla^J\left( ({\boldsymbol I}
-{\boldsymbol Q})\partial_{\boldsymbol l}\hat\eta\right){\boldsymbol l}.
\label{tauring_boundary1}
\end{equation}

Equation \eqref{local_inequality_bulk} gives the entropy production rate per unit volume within the grain. For rigidly deforming grains ${\boldsymbol\sigma}\cdot\nabla{\boldsymbol v} = 0$ and $\div{\boldsymbol v} = 0$. Additionally, if we assume that $\Psi = \hat{\Psi}(\rho)$ then \eqref{local_inequality_bulk} yields $\mu = \partial_\rho \hat{\Psi}$ and ${\boldsymbol j}\cdot\nabla\mu\leq 0$. 
Equation \eqref{local_inequality_gb1} contains the dissipation rate per unit 
area of the GB, with contribution from boundary migration, relative translation of grains at the boundary, GB diffusion, and misorientation change.  The inequality therein forms a basis for postulating kinetic relations for coupled GB motion, as is done in Section \ref{examples_3D_kinetics}. It can also be a starting point for motivating kinetic relations for motion of incoherent phase boundaries with diffusion and curvature dependent boundary energy \cite{cermelli2, gurtin2}, as well as for a variety of physical phenomena involving coherent interfaces \cite{fried1}. An analogous inequality, valid for a one dimensional interface in a 2D grain, was derived recently by the authors \cite{basak1}. 

Equation \eqref{local_inequality_junc1} gives the net dissipation rate per unit length of the junction curve, with contribution due to motion of the curve, diffusion along it, and evolution of orientation of the intersecting boundaries. A comment is in order regarding the contribution due to the latter, represented by the last term on the L.H.S. of the inequality. It is evident from \eqref{gb_energy_L} and \eqref{c_dot_tau_3D} that this term is linear in scalar parameters ($\epsilon_1$ and $\epsilon_2$) which appear in the curvature dependent part of the GB energy. The scalar parameters are usually infinitesimally small, ensuring that curvature dependent part of the energy is significant only at the corners \cite{gurtin2}. The curvatures of intersecting boundaries, as they approach the junction, are finite and hence $|{\boldsymbol\tau}_i|$ are small. We can therefore ignore the last term on the L.H.S. of the inequality \eqref{local_inequality_junc1} within the present analysis. Secondly, in the context of GB dynamics, we assume the density field to remain bounded at the junction curve and the velocity in each grain to be resulting only a simple rigid body motion (hence no strains in the grain). With these assumptions, and keeping in mind the weak singularity condition \eqref{blmjun_junc}, we can show that the closed integral terms in \eqref{local_inequality_junc1} vanish in the limiting sense. We can rewrite \eqref{local_inequality_junc1} and \eqref{junction_force_expression_3D} under all these considerations as
\begin{equation}
{\mathfrak D}_J = {\boldsymbol{\mathcal F}}_J\cdot{\boldsymbol q}_p
-h_J (\nabla^J \mu)\cdot{\boldsymbol l} \geq 0  ~\forall{\boldsymbol x}\in J,~\text{where}
\label{local_inequality_junc2}
\end{equation}
\begin{equation}
{\boldsymbol{\mathcal F}}_J=-({\boldsymbol I}-{\boldsymbol Q})\left(\sum_{i=1}^3{\boldsymbol c}_i+{\boldsymbol f}_J
\right),
\label{junction_force_expression_3D2}
\end{equation}
During thermodynamic equilibrium, with junction curve remaining stationary and diffusion absent, ${\boldsymbol{\mathcal F}}_J = {\boldsymbol 0}$  which in the absence of junction energy yields the well known Herring's relation \cite{herring2}, i.e. $\sum_{i=1}^3\left(\gamma_i{\boldsymbol t}_i-\partial_{{\boldsymbol n}_i}\gamma_i\right)
={\boldsymbol 0}$. The above framework can be used to obtain the extension of  Herring's relation in the presence of junction energy and various singular fields (see also \cite{simha2}). Finally, we note that junctions have been previously treated in the framework of continuum thermodynamics but only for intersecting boundaries which are coherent \cite{simha2, mariano1}. The grain boundaries however are in general incoherent and a treatment of coupled GB motion necessarily requires allowance for relative slip at the boundary. The present framework allows for such incoherency and for junctions which are formed at the intersection of such boundaries.

\section{Geometric coupling factor}
\label{coupling_factor_3D_deriv}
During the migration of a tilt or a mixed GB, the adjacent grains undergo a tangential 
motion giving rise to a coupled dynamics (Chapter 14 in \cite{read2}, and \cite{cahn2,molodov1,gorkaya2}). The deformation of the grain in the wake of a moving GB, during coupled motion,  is essentially controlled by the intrinsic edge dislocation content at the GB. Screw dislocations, if 
present, just glide along the GB plane and contribute only to grain sliding without affecting the coupling 
process \cite{gorkaya2}. For a moving planar symmetric tilt GB, whose wake experiences a simple shear deformation, Cahn and coworkers
\cite{cahn2} introduced {\em geometric coupling factor} 
as the ratio of the relative tangential velocity 
(in the absence of viscous sliding) to the GB velocity. For a large misorientation range
of a symmetric tilt 
boundary, containing single array of edge dislocations, the coupling factor (denoted by $\beta$) was calculated 
to be $\beta=2\tan (\theta/2)$, where $\theta$ is the misorientation angle. This was later verified both in
experiments and atomistic simulations \cite{molodov1,cahn3}. In general, however, most of the GBs are mixed, containing multiple sets of edge and screw 
dislocation arrays. If the above definition of the geometric coupling factor
is generalized to an arbitrary GB, the result will be a vector given by 
\begin{equation}
{\boldsymbol \beta} = \frac{ {\boldsymbol P} [\![{\boldsymbol v}]\!]}{V}.
\label{coupling_factor_def_3D}
\end{equation}
We will now use this definition to derive an expression for the coupling factor for three special cases: (all for a planar GB 
${\mathscr S}$ as shown in Figure \ref{general_planar_GB}) i) Symmetric tilt boundary with finite misorientation. Here we will provide an alternate argument to recover the formula obtained earlier by Cahn et al. \cite{cahn2}. ii) Twist GB with small misorientation. We show that the coupling factor for such a boundary is zero. iii) Mixed GB with small misorientation.

\subsection{Symmetric tilt GB} 
\label{symm_tilt_oneset_disl}
\begin{figure}[t!]
\centering
    \includegraphics[width=2.2in, height=1.8in] {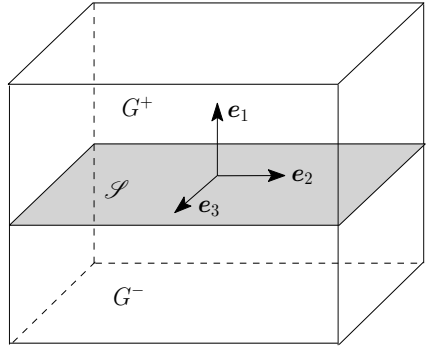}
     \caption{Schematic of a bicrystal with a planar GB.}
     \label{general_planar_GB}
\end{figure}
Let ${\boldsymbol F}$ be the total deformation gradient 
of the grains with respect to a fixed reference configuration. Compatibility at the boundary requires $[\![{\boldsymbol F}]\!]={\boldsymbol a}\otimes{\boldsymbol n}_r$ and $[\![{\boldsymbol v}]\!]=-V_r{\boldsymbol a}$,
where ${\boldsymbol a}$ is an arbitrary vector, while ${\boldsymbol n}_r$ and $V_r$, 
respectively, are the normal vector and the normal velocity of the GB in the reference configuration.
Without loss of generality, we can assume that ${\boldsymbol F}^+={\boldsymbol I}$ and 
${\boldsymbol v}^+={\boldsymbol 0}$. Consequently, ${\boldsymbol n}_r={\boldsymbol n}$ and $V_r=V$ \cite{gupta1}. On the other hand the multiplicative decomposition of ${\boldsymbol F}$, under the present assumption of elastically rigid grains, takes the form ${\boldsymbol F}={\boldsymbol R} {\boldsymbol F}^p$ \cite{gupta1}, where ${\boldsymbol R}$ is the lattice rotation tensor and ${\boldsymbol F}^p$ is the plastic deformation gradient. 
If we assume the plastic deformation to be isochoric ($\text{det} 
{\boldsymbol F}^p=1$), and that ${\boldsymbol F}^{p+} = {\boldsymbol I}$, then the above considerations lead to ${\boldsymbol F}^-={\boldsymbol I}+{\boldsymbol \beta}\otimes{\boldsymbol n}$,
where ${\boldsymbol\beta}=- {\boldsymbol P}{\boldsymbol a}$ is a tangential vector (the superscript `-' will be suppressed hereafter). 
The total Burgers vector ${\boldsymbol B}$ of all the GB dislocations cut by a unit vector
${\boldsymbol p}$ lying on the GB plane is given by the Frank-Bilby equation \cite{gupta1}
\begin{equation}
{\boldsymbol B} = ({\boldsymbol I} - {\boldsymbol R}^T){\boldsymbol p}=({\boldsymbol I} - {\boldsymbol F}^p){\boldsymbol p}.
\label{frank_bibly_equation}
\end{equation}
Assuming all the edge dislocations at the GB to glide in a single slip direction, we can write the resulting plastic distortion rate  as $\dot{\boldsymbol F}^p({\boldsymbol F}^p)^{-1}= \dot\zeta {\boldsymbol s}\otimes{\boldsymbol m}$ (cf. Chapter $106$ in \cite{gurtin_cont_mech}),
 where $\dot\zeta$,  ${\boldsymbol s}$, and ${\boldsymbol m}$ stand for slip rate, unit slip vector, and 
unit normal to the slip plane, respectively (${\boldsymbol s}$ and ${\boldsymbol m}$ are mutually perpendicular). With initial values of $\zeta$ and ${\boldsymbol F}^p$ as $0$ and ${\boldsymbol I}$ respectively, time 
integration of the evolution equation yields ${\boldsymbol F}^p = \exp(\zeta(t) {\boldsymbol s}\otimes{\boldsymbol m}) 
({\boldsymbol F}^p|_{t=0})={\boldsymbol I}+\zeta {\boldsymbol s}\otimes{\boldsymbol m}$. If the orientation of grain $G^-$ is related to that of $G^+$ by an anticlockwise 
rotation of angle $\theta_3$ about ${\boldsymbol e}_3$-axis (see Figure \ref{general_planar_GB}) then we can write ${\boldsymbol R}=\cos\theta_3 ({\boldsymbol e}_1\otimes{\boldsymbol e}_1+{\boldsymbol e}_2\otimes
{\boldsymbol e}_2)+\sin\theta_3 ({\boldsymbol e}_2\otimes{\boldsymbol e}_1-{\boldsymbol e}_1
\otimes{\boldsymbol e}_2)+{\boldsymbol e}_3\otimes{\boldsymbol e}_3$. Using this in 
\eqref{frank_bibly_equation}$_1$ for ${\boldsymbol p}={\boldsymbol e}_2$, and recalling that
${\boldsymbol B}=|{\boldsymbol B}|{\boldsymbol s}$, we obtain $|{\boldsymbol B}|=2\sin(\theta_3 /2)$ and
${\boldsymbol s} = -\cos(\theta_3/2) {\boldsymbol e}_1+\sin(\theta_3/2) {\boldsymbol e}_2$ (and hence ${\boldsymbol m} = -\sin(\theta_3/2) {\boldsymbol e}_1-\cos(\theta_3/2) {\boldsymbol e}_2$). Finally, with the help of expressions derived above for total and plastic deformation gradients, we can obtain $\zeta= -{|{\boldsymbol B}|}/({{\boldsymbol m}\cdot{\boldsymbol e}_2})$ and consequently,
\begin{equation}
{\boldsymbol\beta}= 2\tan(\theta_3/2) {\boldsymbol e}_2.
\label{beta_vector_tiltGB1}
\end{equation} 
This expression for the coupling factor was derived earlier by Cahn et al. \cite{cahn2}. For small misorientation angle the coupling factor takes a simple form ${\boldsymbol\beta}= \theta_3 {\boldsymbol e}_2$ (Chapter $14$ in \cite{read2}). 

\noindent \textit{A general tilt boundary with small misorientation}: Restricting ourselves to small misorientation, we consider a tilt GB such that ${\boldsymbol R}={\boldsymbol  I}+\theta_2({\boldsymbol e}_2\times)+\theta_3({\boldsymbol e}_3\times)$, where $({\boldsymbol e}\times)$ represents a skew tensor with components given by  $({\boldsymbol e}\times)_{jk}=\varepsilon_{jlk}e_l$ (here $\varepsilon_{jlk}$ is the permutation symbol). In other words, grain $G^-$ is obtained by rotating the reference grain $G^+$ anticlockwise about 
${\boldsymbol e}_2$ and ${\boldsymbol e}_3$ by small angles $\theta_2$ and $\theta_3$, respectively. The dislocation density tensor at the GB, defined as ${\boldsymbol \alpha}=({\boldsymbol  I} - {\boldsymbol F}^p)({\boldsymbol n}\times)=({\boldsymbol  I} - {\boldsymbol R}^T)({\boldsymbol n}\times)$ \cite{gurtin5, gupta1}, takes the form
\begin{equation}
{\boldsymbol \alpha}=\theta_2{\boldsymbol e}_1
\otimes{\boldsymbol e}_2 + \theta_3{\boldsymbol e}_1\otimes{\boldsymbol e}_3.
 \label{disl_density_low_GB_many_disl1}
\end{equation}
This represents two arrays of edge dislocations having line direction along ${\boldsymbol e}_2$ and ${\boldsymbol e}_3$, and slip direction ${\boldsymbol e}_1$, with densities $\theta_2$ and $\theta_3$, respectively. To calculate the geometric coupling factor we exploit linearity in extending the above result for a symmetric tilt boundary to the present situation to obtain
\begin{equation}
{\boldsymbol\beta}= \theta_3\,{\boldsymbol e}_2-\theta_2\,{\boldsymbol e}_3.
 \label{low_GB_planar_two_disl1}
\end{equation}
The coupling factor therefore has contributions from both the arrays of edge dislocation. 
 
\subsection{Twist GB with small misorientation}
\label{twist_gb} 
 
We now consider a twist GB with small misorientation such that the grain $G^-$ is rotated by an anticlockwise angle $\theta_1$, about ${\boldsymbol e}_1$-axis, with respect to grain $G^+$. For small angle we can write ${\boldsymbol R}={\boldsymbol  I}+\theta_1({\boldsymbol e}_1\times)$. As a result
\begin{equation}
{\boldsymbol \alpha}= -\theta_1({\boldsymbol e}_2\otimes{\boldsymbol e}_2+{\boldsymbol e}_3
\otimes{\boldsymbol e}_3),
 \label{disl_density_low_GB_twist}
\end{equation}
which represents two arrays of screw dislocations (both with density $\theta_1$) with line directions parallel to ${\boldsymbol e}_2$ and ${\boldsymbol e}_3$. With this in mind we assume the plastic deformation gradient as  ${\boldsymbol F}^p ={\boldsymbol I}+\zeta ({\boldsymbol s}_1\otimes{\boldsymbol m}_1 + {\boldsymbol s}_2\otimes{\boldsymbol m}_2)$, where we have considered two mutually-orthogonal slip systems with equal slip magnitude, such that $|{\boldsymbol s}_1| = |{\boldsymbol s}_2|  = |{\boldsymbol m}_1| =|{\boldsymbol m}_2| =1$ and ${\boldsymbol s}_1 \cdot {\boldsymbol s}_2 = {\boldsymbol s}_1 \cdot {\boldsymbol m}_1 = {\boldsymbol s}_2 \cdot {\boldsymbol m}_2 = {\boldsymbol m}_1 \cdot {\boldsymbol m}_2 = 0$. On the other hand, the total deformation gradient is of the form considered above, i.e. ${\boldsymbol F}={\boldsymbol I}+{\boldsymbol \beta}\otimes{\boldsymbol e}_1$. With the assumption of small deformation and small misorientation, the multiplicative decomposition of the deformation gradient becomes an additive decomposition so as to yield the following for the case at hand:
\begin{equation}
{\boldsymbol \beta}\otimes{\boldsymbol e}_1=\theta_1({\boldsymbol e}_1\times)+\zeta ({\boldsymbol s}_1\otimes{\boldsymbol m}_1 + {\boldsymbol s}_2\otimes{\boldsymbol m}_2).
 \label{total_distorsion_infinit_decompos}
\end{equation}
Projecting this onto ${\boldsymbol e}_1$, ${\boldsymbol e}_2$, and ${\boldsymbol e}_3$ we obtain
\begin{eqnarray}
&& {\boldsymbol \beta}=\zeta \left({\boldsymbol s}_1 ({\boldsymbol m}_1 \cdot {\boldsymbol e}_1) + {\boldsymbol s}_2 ({\boldsymbol m}_2 \cdot {\boldsymbol e}_1) \right), \label{beta_twist}\\
&& 0= \theta_1 {\boldsymbol e}_3 + \zeta \left({\boldsymbol s}_1 ({\boldsymbol m}_1 \cdot {\boldsymbol e}_2) + {\boldsymbol s}_2 ({\boldsymbol m}_2 \cdot {\boldsymbol e}_2) \right),~\text{and}\\
&& 0= -\theta_1 {\boldsymbol e}_2 + \zeta \left({\boldsymbol s}_1 ({\boldsymbol m}_1 \cdot {\boldsymbol e}_3) + {\boldsymbol s}_2 ({\boldsymbol m}_2 \cdot {\boldsymbol e}_3) \right),
\end{eqnarray}
respectively. After some manipulations, the latter two equations yield ${\boldsymbol s}_1 \cdot {\boldsymbol e}_1 = {\boldsymbol s}_2 \cdot {\boldsymbol e}_1 =0$, ${\boldsymbol m}_1 = {\boldsymbol s}_2$, and ${\boldsymbol m}_2 = {\boldsymbol s}_1$. These results, when substituted into \eqref{beta_twist}, immediately furnish ${\boldsymbol \beta}={\boldsymbol 0}$. We have therefore shown that the geometric coupling factor for a twist boundary (with small misorientation) is zero, hence confirming the qualitative arguments provided in \cite{cahn1, cahn2, gorkaya2}.

\subsection{A cubic grain embedded in a large grain}
As an application of the results obtained in the previous two subsections we now consider an example where a 
cubic grain (whose edges are aligned with directions ${\boldsymbol e}_1$, ${\boldsymbol e}_2$, and ${\boldsymbol e}_3$) is embedded inside another grain such that the (infinitesimal) misorientation between 
them is given by 
\begin{equation}
{\boldsymbol R}={\boldsymbol I} + \theta_1({\boldsymbol e}_1\times)+\theta_2({\boldsymbol e}_2
\times)+\theta_3({\boldsymbol e}_3\times).
 \label{orinetation_cube_low_GB_many_disl2}
\end{equation}
The surface dislocation density tensor for the GB with 
normal ${\boldsymbol e}_1$ can be calculated as
\begin{equation}
{\boldsymbol \alpha}=-\theta_1({\boldsymbol e}_2\otimes{\boldsymbol e}_2+{\boldsymbol e}_3
\otimes{\boldsymbol e}_3)+\theta_2{\boldsymbol e}_1
\otimes{\boldsymbol e}_2+\theta_3{\boldsymbol e}_1\otimes{\boldsymbol e}_3,
 \label{disl_density_low_GB_many_disl2}
\end{equation}
which is the sum of densities given in \eqref{disl_density_low_GB_many_disl1} and \eqref{disl_density_low_GB_twist}; the GB is of a mixed type consisting of two mutually perpendicular 
sets of edge dislocations (with densities $\theta_2$ and $\theta_3$) and two mutually perpendicular sets of screw dislocations (both with densities $\theta_1$). 
The dislocation content at other boundaries can be obtained in a similar manner. In evaluating the geometric coupling factor associated with the boundary with normal ${\boldsymbol e}_1$, we exploit linearity in our arguments (due to small misorientation) to combine the results obtained above for tilt and twist boundaries to write
\begin{equation}
{\boldsymbol\beta}_{{\boldsymbol e}_1}=\theta_3\,{\boldsymbol e}_2-\theta_2\,{\boldsymbol e}_3. \label{beta_cube_e1}
\end{equation}
We can similarly calculate the coupling factors for the GBs with normal ${\boldsymbol e}_2$ 
and ${\boldsymbol e}_3$ as
\begin{equation}
{\boldsymbol\beta}_{{\boldsymbol e}_2}=-\theta_3\,{\boldsymbol e}_1
+\theta_1\,{\boldsymbol e}_3,~\text{and}~{\boldsymbol\beta}_{{\boldsymbol e}_3}=\theta_2\,
{\boldsymbol e}_1-\theta_1\,{\boldsymbol e}_2,
\end{equation}
respectively. It is easily verifiable that ${\boldsymbol\beta}_{-{\boldsymbol e}_1}
=-{\boldsymbol\beta}_{{\boldsymbol e}_1}$, etc, where ${\boldsymbol\beta}_{-{\boldsymbol e}_1}$ represents the coupling factor associated with the face with normal $-{\boldsymbol e}_1$.

\section{Kinetic relations}
\label{examples_3D_kinetics}
The governing equations for coupled GB dynamics with junctions can be derived starting from inequalities \eqref{local_inequality_gb1} and \eqref{local_inequality_junc1} by first identifying various dissipative fluxes, and the associated driving forces, and then assuming linear kinetics. Towards this end we consider three crystalline arrangements: (i) bicrystal-I consisting of two rectangular grains joined at a mixed planar 
GB (as in Figure \ref{general_planar_GB}) and subjected to shear stress;
(ii) bicrystal-II with a spherical grain embedded inside a larger grain (see Figure \ref{spherical_3D}); 
and
(iii) tricrystal where a grain $G_1$ is embedded inside a large bicrystal 
made of two rectangular grains $G_2$ and $G_3$ (see Figure \ref{tricrystal_3D}).

\subsection{Bicrystal-I}
\label{bicrystal_stress_mixed_GB}
The first bicrystalline arrangement is as shown in Figure \ref{general_planar_GB} such that the traction on the outer boundaries perpendicular to ${\boldsymbol e}_1$ is $\tau{\boldsymbol e}_2$
and the traction on the outer boundaries perpendicular to ${\boldsymbol e}_2$ is $\tau{\boldsymbol e}_1$. We also assume that the 
grains are rigid, free of defects, and contain negligible stored energy. Moreover, we neglect all kinds 
of atomic diffusion. The GB ${\mathscr S}$ is considered to be of a mixed type, with both tilt and twist components, where the misorientation is given by
\begin{equation}
{\boldsymbol \Theta}={\boldsymbol I} + \theta_1({\boldsymbol e}_1\times)+\theta_3({\boldsymbol e}_3\times),
 \label{bicrystal1_misorientation_tensor}
\end{equation}
for small $\theta_1$ and $\theta_3$; as discussed previously,  
$\theta_1$ and $\theta_3$ determine the twist and the tilt characteristic, respectively, of the GB. Whereas the deformation of grains in the wake of a moving GB, under the external loading considered here, is simple shear for a tilt GB, it is more
complicated if the GB is of mixed type \cite{gorkaya2}. The array of edge dislocations are driven by the Peach-Koehler force to move the GB in normal 
direction while translating the grains parallel to the GB. On the other hand, the  
simultaneous movement of two perpendicular sets of screw dislocation arrays results into a relative rotation of the adjacent grains about the GB normal.
The GB motion, the relative tangential translation, and the grain rotation are in general all coupled to 
each other. 
 
The state of stress throughout the bicrystal is taken as ${\boldsymbol\sigma}=\tau({\boldsymbol e}_2\otimes{\boldsymbol e}_1+{\boldsymbol e}_1
\otimes{\boldsymbol e}_2)$; this clearly satisfies both the equilibrium equations and the traction boundary conditions. Based on the experimental
observations in \cite{molodov1, gorkaya2}, we assume the tilt angle to remain fixed while allowing the 
twist angle to evolve owing to the relative rotation between the grains. The axis of rotation is taken to coincide with ${\boldsymbol e}_1$. Without loss of generality, the grain $G^+$ can be assumed to remain stationary, 
i.e. ${\boldsymbol v}^+={\boldsymbol 0}$, and $G^-$ moving with a velocity
\begin{equation}
{\boldsymbol v}^-=\dot\theta_1{\boldsymbol e}_1\times{\boldsymbol x}+\dot{C}{\boldsymbol e}_2,
 \label{dissipation_ineql_planar_GB1}
\end{equation} 
where $\dot{C}$ is the translational velocity of $G^-$ in the direction of 
${\boldsymbol e}_2$ and ${\boldsymbol x}=x_1{\boldsymbol e}_1+x_2{\boldsymbol e}_2
+x_3{\boldsymbol e}_3$ is the position vector. 
Observing that $[\![v_n]\!]=0$, 
the dissipation inequality \eqref{local_inequality_gb1} reduces to
\begin{equation}
\dot\theta_1 f_\theta+\dot{C}f_c\geq 0,~\forall{\boldsymbol x}\in{\mathscr S}
 \label{dissipation_ineql_planar_GB}
\end{equation} 
where $f_\theta=\displaystyle\left(\tau x_3-{\partial\gamma}/{\partial\theta_1}\right)$ and $f_c = -\tau$.
Considering Onsager's reciprocity theorem \cite{onsager1}, we can obtain the following pair of coupled kinetic relations \cite{basak1}:
\begin{equation}
\dot\theta_1 = {\mathcal S}_d f_\theta + {\mathcal B}{\mathcal S}_d f_c~\text{and}~\dot{C} = {\mathcal B}\dot\theta_1+{\mathcal L} f_c,
 \label{kinetic_laws_planar_GB3}
\end{equation}
where ${\mathcal S}_d\geq 0$ is the sliding coefficient due to the relative rotational motion
(caused by the intrinsic screw dislocation glide along ${\mathscr S}$), 
${\mathcal L}\geq 0$ is the sliding coefficient for the relative translational motion between the grains, 
and ${\mathcal B}$ denotes a coupling between grain rotation and translation. In response to grain translation, the edge dislocation array will cause 
simultaneous GB migration \cite{cahn2}, such that
\begin{equation}
V=-\dot{C}/\beta_2,
 \label{coupling1_planar_GB3}
\end{equation}
where $\beta_2 = \theta_3$ is the geometric coupling factor as calculated in the previous section. Note that the geometric coupling exists only with respect to the translational velocity, as the rotational part in \eqref{dissipation_ineql_planar_GB1} amount to pure sliding. According to 
\eqref{coupling1_planar_GB3} it is the direction of $\dot{C}$ which decides whether the GB will move upwards
or downwards. 

To summarize, we have a coupled system of equations, given by
\eqref{kinetic_laws_planar_GB3} and \eqref{coupling1_planar_GB3}, which should be solved to evaluate the position of the grains and the GB, as well as the misorientation, at any given time instance during the dynamical process. It should be noted that, in a more complicated situation when the driving forces are functions of $x_1$ and $x_2$, 
an initially planar GB will not necessarily remain planar (cf. \cite{gorkaya2}) and GB diffusion will be required to prevent void-formation/interpenetration at the GB. 

\subsection{Bicrystal-II}
\label{examples_junc}
\begin{figure}[t!]
\centering 
\subfigure[]{
  \includegraphics[width=2in, height=1.6in] {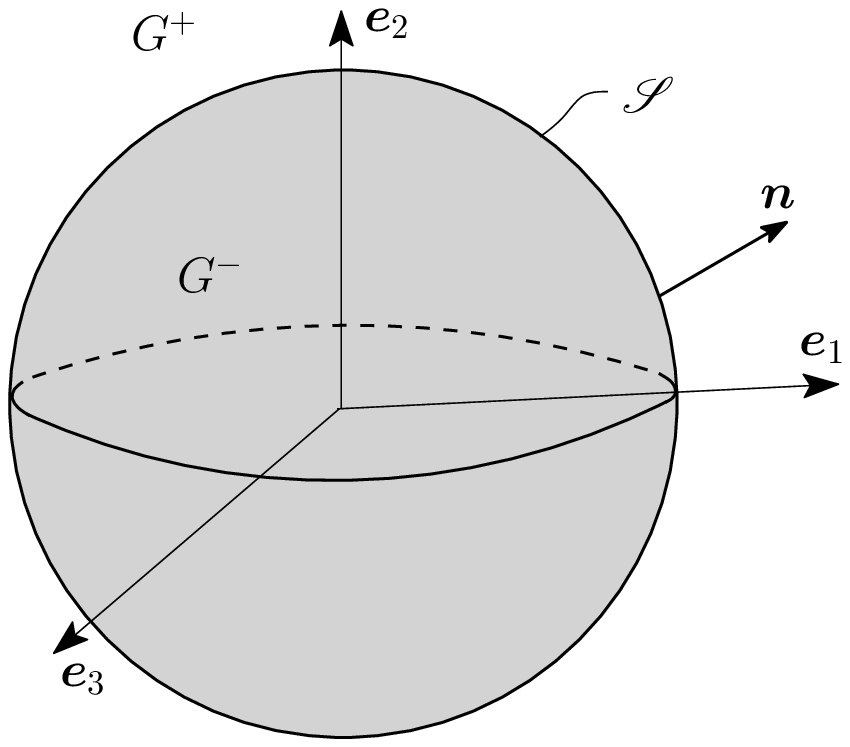}
	\label{spherical_3D}}
\hspace{10mm}
    \subfigure[]{
    \includegraphics[width=2.5in, height=1.7in] {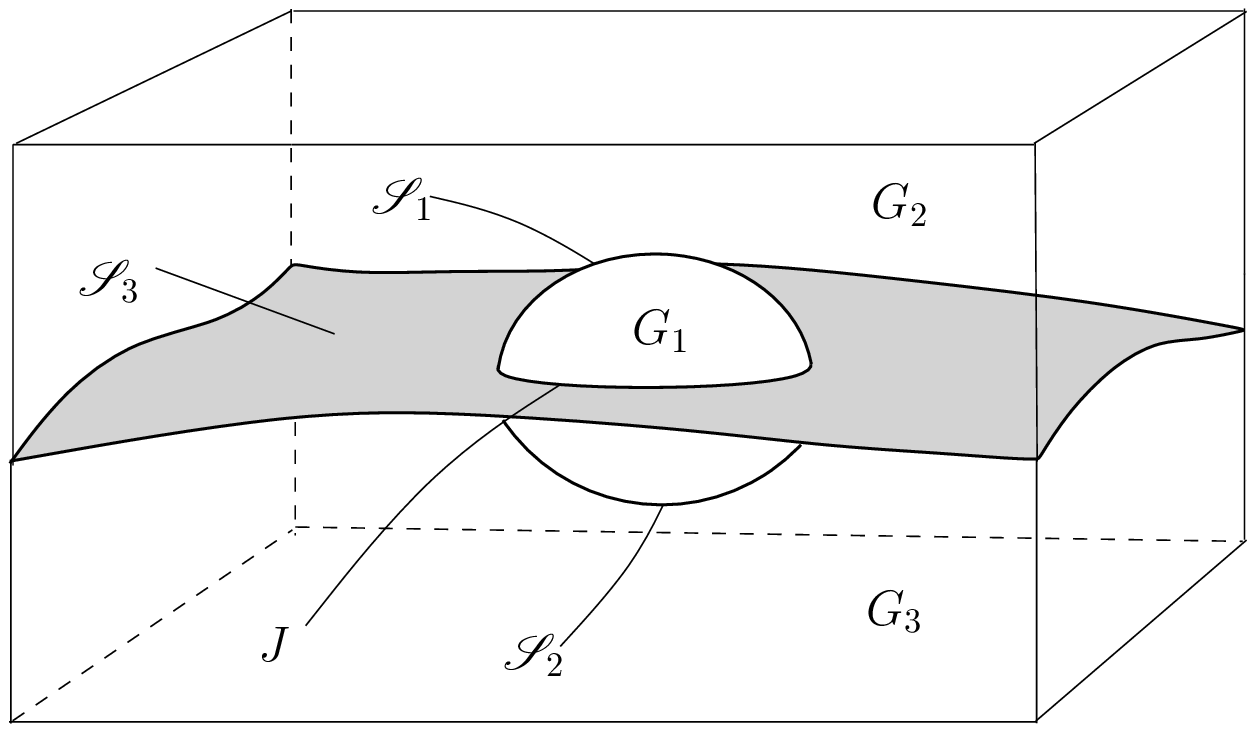}
    \label{tricrystal_3D}}
 \caption{(a) Schematic of bicrystal-II. (b) Schematic of tricrystal. }
\end{figure}

As a second example, we consider a bicrystal with a spherical grain $G^-$ (of radius $R$) embedded within a much larger grain $G^+$, as shown in Figure \ref{spherical_3D}, with misorientation between the grains given by
\begin{equation}
{\boldsymbol \Theta}={\boldsymbol I} + \theta_1 ({\boldsymbol e}_1\times)+\theta_2 ({\boldsymbol e}_2\times)+\theta_3
({\boldsymbol e}_3\times),
 \label{misorientation_skew}
\end{equation}    
i.e. grain $G^-$ has been obtained by rotating it from $G^+$ by small angles $\theta_1$, $\theta_2$, 
and $\theta_3$ about ${\boldsymbol e}_1$, ${\boldsymbol e}_2$, and ${\boldsymbol e}_3$, respectively, where the orthonormal basis vectors $\{{\boldsymbol e}_1,{\boldsymbol e}_2,{\boldsymbol e}_3\}$ form a coordinate frame with origin at the center of the sphere. The GB $\mathscr{S}$ hence is of a mixed type. Without loss of generality we let the outer grain $G^+$ to 
remain fixed and allow the inner grain $G^-$ to rotate (without translating). We assume that the external stress 
is absent, the grains are rigid and free of defects, the free energy of the grains is vanishing, 
and the volumetric diffusion is absent. Let us
consider an orthonormal spherical basis $\{{\boldsymbol e}_R,{\boldsymbol e}_\xi,{\boldsymbol e}_\phi\}$ with origin
at the center of the embedded grain ($\xi$ is the polar angle and $\phi$ is the azimuthal angle).
The GB normal ${\boldsymbol n}$ points into $G^+$, hence
${\boldsymbol n}={\boldsymbol e}_R$. The angular velocity of $G^-$ (axial vector of $\dot{\boldsymbol\Theta}$) is given by 
\begin{equation}
{\boldsymbol w}=\dot\theta_1\,{\boldsymbol e}_1+\dot\theta_2\,{\boldsymbol e}_2+\dot\theta_3\,{\boldsymbol e}_3.
 \label{angular_velocity_3D_skew}
\end{equation}
Neglecting rigid body translation, the velocity of the inner grain can be written as 
${\boldsymbol v}^-={\boldsymbol w}\times{\boldsymbol x}$, where ${\boldsymbol x}$ is the position vector. Recalling that ${\boldsymbol v}^+={\boldsymbol 0}$  we obtain
\begin{equation}
[\![v_n]\!]=v_n^+= 0 ~\text{and}~
{\boldsymbol V}_t={\boldsymbol P}({\boldsymbol v}^+-{\boldsymbol v}^-)=-{\boldsymbol v}^-,
 \label{velocity_spherical_tang}
\end{equation} 
where
${\boldsymbol V}_t$ represents the relative tangential velocity of $G^-$ with respect to $G^+$ at the GB.

The spherical GB rotates and shrinks without any shape change. This does not require any shape accommodation mechanism such as GB diffusion. We will therefore consider ${\boldsymbol h}={\boldsymbol 0}$. 
Assuming isotropic GB energy and defining ${\boldsymbol\theta}=\theta_1{\boldsymbol e}_1
+\theta_2{\boldsymbol e}_2+\theta_3{\boldsymbol e}_3$ (the axial vector of
${\boldsymbol\Theta} - {\boldsymbol I}$) the dissipation inequality, given by \eqref{local_inequality_gb1}, reduces to 
 \begin{equation}
Vf_n+{\boldsymbol w}\cdot\tilde{\boldsymbol f}_t\geq 0,~\text{where}
\label{local_dissi_inequality_shpere1}
\end{equation}  
\begin{equation}
f_n=\gamma\kappa~\text{and}~
\tilde{\boldsymbol f}_t=-\partial_{\boldsymbol\theta}\gamma.
\label{tang_force_shpere}
\end{equation} 
Invoking the Onsager reciprocity theorem 
\cite{onsager1}, we can postulate the following linear kinetic relations based on \eqref{local_dissi_inequality_shpere1} \cite{basak1}:
\begin{equation}
V = {\mathcal M}\,f_n+{\mathcal M}\tilde{\boldsymbol\beta}\cdot\tilde{\boldsymbol f}_t~\text{and}
 \label{kinetic_relation_Un2}
\end{equation}
\begin{equation}
{\boldsymbol w}=\tilde{\boldsymbol\beta}\,V+\tilde{\boldsymbol{\mathcal S}}
\tilde{\boldsymbol f}_t,
 \label{kinetic_relation_Utan2}    
\end{equation}
where ${\mathcal M}>0$ is the GB mobility, $\tilde{\boldsymbol\beta}$ is the coupling factor between
rotational speed and normal GB velocity, and $\tilde{\boldsymbol{\mathcal S}}$ is the symmetric positive 
semi-definite sliding coefficient. Both the viscous effect and the twist characteristic of the GB are 
expected to contribute to the net sliding. To understand the physical meaning of the coefficients  $\tilde{\boldsymbol\beta}$ 
 and $\tilde{\boldsymbol{\mathcal S}}$, we take a cross-product of \eqref{kinetic_relation_Utan2} 
 with ${\boldsymbol x}$ to obtain
\begin{equation}
{\boldsymbol V}_t={\boldsymbol\beta}\,V+{\boldsymbol{\mathcal S}}{\boldsymbol f}_t,~\text{where}
 \label{inner_grain_velocity}
\end{equation} 
\begin{equation}
{\boldsymbol\beta}={\boldsymbol x} \times \tilde{\boldsymbol\beta}, ~
{\boldsymbol{\mathcal S}}{\boldsymbol f}_t={\boldsymbol x} \times \tilde{\boldsymbol{\mathcal S}}\tilde{\boldsymbol f}_t,~\text{and}~{\boldsymbol f}_t=(1/R) {\boldsymbol e}_R \times \tilde{\boldsymbol f}_t.
 \label{kinetic_ciefficients_3D}
\end{equation}
Equation \eqref{kinetic_ciefficients_3D}$_1$ relates $\tilde{\boldsymbol\beta}$ to the 
the geometric coupling factor ${\boldsymbol\beta}$ introduced in the previous section, while \eqref{kinetic_ciefficients_3D}$_2$ relates $\tilde{\boldsymbol{\mathcal S}}$
to the sliding 
coefficient ${\boldsymbol{\mathcal S}}$; the relation \eqref{kinetic_ciefficients_3D}$_3$ is motivated from the 2D counterpart of the present discussion \cite{basak1}. Next, we specialize these kinetic equations for a spherical GB, under various additional assumptions, and present analytical solutions wherever possible. When reduced to a 2D setting, the derived relations will be identical to those obtained for a circular GB in \cite{cahn1} and \cite{basak1}.

\noindent \textit{GB migration}:
When both geometric coupling and GB sliding are negligible, kinetic equations 
\eqref{kinetic_relation_Un2} and \eqref{kinetic_relation_Utan2} simplify to the well-known
equations for curvature driven GB migration: $V={\mathcal M}\gamma\kappa$ and 
${\boldsymbol w}={\boldsymbol 0}$. Using $V=\dot{R}$ and $\kappa=-1/R$ (for a spherical GB)
in the governing equations we obtain the following solutions:  
\begin{equation}
R(t)=\sqrt{R_0^2-2{\mathcal M}\gamma t}~\text{and}~
{\boldsymbol\theta}(t)={\boldsymbol\theta}_0,
 \label{sphere_GB_migration1}
\end{equation}
where  $R_0$ is the initial GB radius and ${\boldsymbol \theta}_0$ is the initial misorientation. Both ${\mathcal M}$ and $\gamma$ have been treated as constants.

\noindent \textit{Coupled motion without GB sliding}: At temperatures far below the melting point, GB viscous sliding 
becomes negligible and geometric coupling plays the dominant role in grain rotation \cite{cahn2}.  
The kinetic relations \eqref{kinetic_relation_Un2} and 
\eqref{kinetic_relation_Utan2} can then be assumed to take the form
\begin{equation}
V={\mathcal M}\gamma\kappa+{\mathcal M}\tilde{\boldsymbol\beta}\cdot\tilde{\boldsymbol f}_t
~\text{and}~{\boldsymbol w}=\tilde{\boldsymbol \beta}V,
 \label{sphere_GB_migration2}
\end{equation}
respectively. For a spherical GB (with $V=\dot{R}$ and ${\boldsymbol\beta}=({\boldsymbol\Theta}- {\boldsymbol I}){\boldsymbol e}_R$), \eqref{kinetic_ciefficients_3D}$_1$ and a cross product of \eqref{sphere_GB_migration2}$_2$ with ${\boldsymbol x}
=R\,{\boldsymbol e}_R$ furnishes
\begin{equation}
{\boldsymbol w}\times{\boldsymbol e}_R=-\frac{\dot{R}}{R}{\boldsymbol\theta}\times
{\boldsymbol e}_R,
 \label{sphere_GB_migration3}
\end{equation}
which on time integration gives
\begin{equation}
R(t){\boldsymbol\theta}(t) = R_0 {\boldsymbol\theta}_0
 \label{sphere_GB_migration4}
\end{equation}
for all $\xi$ and $\phi$. Note that we have used the same geometric coupling factor 
associated with a planar GB for analysing the kinetics of a curved GB; this assumption is motivated from the 2D atomistic studies where the geometric coupling factor was observed to be 
same for planar and curved GBs (cf. \cite{trautt3} and the references therein). Before integrating \eqref{sphere_GB_migration2}$_1$ we consider the  
following isotropic energy proposed by Read (see Section 12.7 in \cite{read2}):
\begin{equation}
\gamma=\gamma_0|{\boldsymbol\theta}|(A_c-\ln|{\boldsymbol\theta}|),
 \label{read_shockley_general_GB}
\end{equation}
where $\gamma_0$ is a constant depending on the material properties, the Burgers vector,
and the spacing between the dislocations; $A_c$ is a constant which depends 
on the energy of atomic misfit at the dislocation core. Differentiating 
\eqref{read_shockley_general_GB} with respect to ${\boldsymbol\theta}$ we obtain the torque
\begin{equation}
\tilde{\boldsymbol f}_t=-\partial_{\boldsymbol\theta}\gamma=-\gamma_0\frac{{\boldsymbol\theta}}
{|{\boldsymbol\theta}|}(A_c-1-\ln|{\boldsymbol\theta}|).
 \label{read_shockley_general_GB1}
\end{equation}
Substituting this into \eqref{sphere_GB_migration2}$_1$, and restricting it to a spherical GB (for which $V=\dot{R}$, $\kappa = -1/R$, and $\tilde{\boldsymbol\beta} = -{\boldsymbol \theta}/R$),  we obtain
$\dot{R} =-{{\mathcal M}\gamma_0|{\boldsymbol\theta}|}/{R}$. Combining it with  
\eqref{sphere_GB_migration4} and subsequently integrating yields 
\begin{equation}
R(t)= (R_0^3-3{\mathcal M}\gamma_0R_0|{\boldsymbol\theta}_0|t)^{1/3}~\text{and}
~
{\boldsymbol\theta}(t)= R_0(R_0^3-3{\mathcal M}\gamma_0R_0|{\boldsymbol\theta}_0|t)^{-1/3}
{\boldsymbol\theta}_0,
 \label{sphere_GB_migration5}
\end{equation}
where ${\mathcal M}$ has been treated as a constant.

\noindent \textit{Coupled motion without geometric coupling}: At temperatures close to the melting point, viscous sliding at the GB
dominates over the geometric coupling to govern grain rotation \cite{cahn2}.
The kinetic relations \eqref{kinetic_relation_Un2} and \eqref{kinetic_relation_Utan2} in such case
reduce down to   
\begin{equation}
V= {\mathcal M}\gamma\kappa,~ \text{and}~
\dot{\boldsymbol\theta}=\tilde{\boldsymbol{\mathcal S}}\tilde{\boldsymbol f}_t,
 \label{sphere_GB_migration6}
\end{equation}
respectively. 
Solving these equations analytically for $R$ and ${\boldsymbol\theta}$ is challenging;  we will attempt to derive a useful implicit relation between them. If we assume $\tilde{\boldsymbol{\mathcal S}}$ and 
${\boldsymbol{\mathcal S}}$ to be of the form $\tilde{\boldsymbol{\mathcal S}}=
\tilde{\mathcal S}{\boldsymbol P}$ and ${\boldsymbol{\mathcal S}}={\mathcal S}{\boldsymbol I}$, respectively, then \eqref{kinetic_ciefficients_3D}$_2$ requires $\tilde{\mathcal S} 
= {\mathcal S}/R^2$. Consequently, for a spherical GB,  equations \eqref{sphere_GB_migration6} can be manipulated to obtain
\begin{equation}
\frac{\dot{\boldsymbol\theta}}{\dot{R}}=\frac{{\mathcal S}}{\mathcal M\,R}\frac{
{\boldsymbol\theta}\,
(A_c-1-\ln|{\boldsymbol\theta}|)}{|{\boldsymbol\theta}|^2(A_c-\ln|{\boldsymbol\theta}|)},
 \label{sphere_GB_migration7}
\end{equation}
where we have also used \eqref{read_shockley_general_GB} and 
\eqref{read_shockley_general_GB1}.  This is a non-linear equation in ${\boldsymbol\theta}$.
Taking a dot product with ${\boldsymbol\theta}$ on both sides of the equation,
and then integrating the result, we obtain the following implicit relation between $R$ and ${\boldsymbol\theta}$:
\begin{equation}
R({\boldsymbol\theta})=R_0 \exp\left(\frac{\mathcal M}{{\mathcal S}}\left\{|{\boldsymbol\theta}|^2-|
{\boldsymbol\theta}_0|^2+2 e^{2(A_c-1)} [E(1,u_0)-E(1,u)]\right\}\right),
 \label{sphere_GB_migration8}
\end{equation}
where $u=2(A_c-1-\ln|{\boldsymbol\theta}|)$, $u_0=2(A_c-1-\ln|{\boldsymbol\theta}_0|)$, and 
$E(n,y)=\int_y^\infty\frac{e^{-u}}{y^{1-n}u^n}du$ (the exponential integral); ${\mathcal M}$
and ${\mathcal S}$ have been considered to be constants.

\noindent \textit{Fully coupled motion}: We finally consider the situation when
both geometrical coupling and GB sliding will contribute comparably to the evolution of grain rotation.
Assuming $\tilde{\boldsymbol{\mathcal S}}$ to be invertible, and 
eliminating $\tilde{\boldsymbol f}_t$ between \eqref{kinetic_relation_Un2} and 
\eqref{kinetic_relation_Utan2}, we derive for a spherical GB ($\tilde{\boldsymbol \beta} = -{\boldsymbol \theta}/R$, $V = \dot{R}$, and $\kappa = -1/R$)
\begin{equation}
\dot{R} =-\frac{R}{R^2+{\mathcal M}
{\boldsymbol\theta}\cdot\tilde{\boldsymbol{\mathcal S}}^{-1}{\boldsymbol\theta}}
\left({\mathcal M \gamma}+{\mathcal M}
\tilde{\boldsymbol{\mathcal S}}^{-1}{\boldsymbol\theta}\cdot{\boldsymbol w}\right).
 \label{kinetic_relation_Un3}
\end{equation}
An expression for ${\boldsymbol w}$ can be obtained by substituting \eqref{kinetic_relation_Un2} in 
\eqref{kinetic_relation_Utan2}:
\begin{equation}
{\boldsymbol w} = \frac{{\mathcal M}\gamma{\boldsymbol\theta}}{R^2}+
\left(\tilde{\boldsymbol{\mathcal S}}+\frac{\mathcal M}{R^2}{\boldsymbol\theta}\otimes
{\boldsymbol\theta}\right)\tilde{\boldsymbol f}_t.
 \label{kinetic_relation_Utan3}
\end{equation}

\subsection{Tricrystal}
\label{tricrystal_closed}
We consider a tricrystal, as shown in Figure \ref{tricrystal_3D}, where grain $G_1$
is embedded inside a larger bicrystal made of two rectangular grains $G_2$ and $G_3$. The tricrystal is subjected to external stress. The configuration has three GBs ${\mathscr S}_i$, with unit normals denoted by ${\boldsymbol n}_i$ ($i=1,2,3$), and a closed junction curve $J$. The normals are chosen 
such that both ${\boldsymbol n}_1$ and ${\boldsymbol n}_2$ point into $G_1$ whereas ${\boldsymbol n}_3$ points 
into $G_2$. In deriving the kinetic laws we assume the following: (i) the grains are rigid, defect free, and have a vanishing 
stored energy; (ii) volumetric diffusion is ignored;  (iii) the shape accommodation required 
for preventing void-formation/interpenetration at various GBs is accomplished 
by allowing for diffusion along the GBs; (iv) diffusion along the junction is negligible; and (v) 
the magnitude of applied stresses are small enough so that elastic 
and plastic deformation of the grains can be neglected. Under the combined 
effects of GB capillary force and the applied stress field the GBs will migrate, grain $G_1$ 
will rotate and translate (as a rigid body), grains $G_2$ and $G_3$ will translate rigidly relative to each other, and 
the junction $J$ will move in space, all coupled to each other. 
  
In the absence of intra-granular defects the orientation field associated with grain $G_i$, denoted by ${\boldsymbol R}_i$,
will be homogeneous throughout the grain. We define the misorientation tensor for 
the respective GBs as 
${\boldsymbol\Theta}_1={\boldsymbol R}_1^T{\boldsymbol R}_{2}$, ${\boldsymbol\Theta}_2=
{\boldsymbol R}_1^T{\boldsymbol R}_{3}$, and ${\boldsymbol\Theta}_3={\boldsymbol R}_2^T
{\boldsymbol R}_{3}$. We assume grains $G_2$ and $G_3$ to be non-rotating, thereby fixing their orientations once for all. As a result 
\begin{equation}
\dot{\boldsymbol\Theta}_1{\boldsymbol\Theta}_1^T=
\dot{\boldsymbol\Theta}_2{\boldsymbol\Theta}_2^T=-\dot{\boldsymbol R}_1{\boldsymbol R}_1^T. 
\label{grain_misorientation_rate_tricrystal} 
\end{equation}
On the other hand, the velocities of rigidly deforming grains can be written as 
\begin{equation}
{\boldsymbol v}_1={\boldsymbol w}\times{\boldsymbol x}+\dot{\boldsymbol C}_1, ~ 
{\boldsymbol v}_2=\dot{\boldsymbol C}_2, ~{\boldsymbol v}_3={\boldsymbol 0},
\label{particle_velocity_grains}
\end{equation}
where ${\boldsymbol w}$ is the angular velocity of $G_1$ (the axial vector of 
$\dot{\boldsymbol R}_1{\boldsymbol R}_1^T$), ${\boldsymbol x}$ is the position vector, and ${\boldsymbol C}_i$ is the rigid translation of $G_i$ (grain $G_3$ has been assumed to remain fixed). We define the relative translation velocity of adjacent grains at the respective GBs as
$\dot{\boldsymbol{\mathcal C}}_1=\dot{\boldsymbol C}_1-\dot{\boldsymbol C}_2$, $\dot{\boldsymbol{\mathcal C}}_2
=\dot{\boldsymbol C}_1$, and $\dot{\boldsymbol{\mathcal C}}_3=\dot{\boldsymbol C}_2$. Before we substitute these in the mass balance relations, we would additionally assume that the translational velocity of the embedded grain to be much smaller than its rotational velocity. We can justify this on the basis of the atomistic simulation results which do not show any significant translation in the absence of external stress \cite{trautt3}. The small amplitude of external stress considered here would therefore cause only small relative translation. Keeping this in mind and 
using \eqref{particle_velocity_grains} in \eqref{boms_junc}, along with negligible diffusional fluxes in the grain, we derive
\begin{eqnarray}
&& \div^S{\boldsymbol h}_a = -\rho{\boldsymbol x}\times{\boldsymbol n}_a
\cdot{\boldsymbol w}~\text{for}~a=1,2, ~\text{and}
\nonumber
\\
&& \div^S{\boldsymbol h}_3=-\rho\dot{\boldsymbol{\mathcal C}}_3\cdot{\boldsymbol n}_3.
\label{mass_balance_rewrite1} 
\end{eqnarray}
These relations can be integrated and then combined with 
\eqref{bomjun_junc} (where ${\boldsymbol j} = {\boldsymbol 0}$, $h_J=0$, and $\rho$, ${\boldsymbol v}$ are non-singular at the junction) to obtain 
\begin{equation}
{\boldsymbol h}_a={\boldsymbol A}_a{\boldsymbol w}~\text{and}
~{\boldsymbol h}_3={\boldsymbol A}_3\dot{\boldsymbol{\mathcal C}}_3,
\label{diffusion_flux_tricrystal_3D}
\end{equation}
where ${\boldsymbol A}_i$ is a second order tensor which depends on the geometry of ${\mathscr S}_i$ (see \cite{basak2} for a similar 
calculation for a 2D tricrystal). We also relate these fluxes to the chemical potential by assuming the Fick's law for superficial diffusion \cite{gurtin2}
\begin{equation}
{\boldsymbol h}_i=-{\boldsymbol D}_i\nabla^S\mu ~\text{for}~i=1,2,3,
\label{ficks_law_3D}
\end{equation}
where ${\boldsymbol D}_i$ is the (symmetric and tangential) diffusivity tensor along ${\mathscr S}_i$. 

For the present case, the dissipation inequality in the bulk \eqref{local_inequality_bulk} is trivially satisfied. The dissipation inequalities at the GBs, given by \eqref{local_inequality_gb1}, are however non-trivial and will be used in the following to derive the kinetic equations. Substituting \eqref{grain_misorientation_rate_tricrystal}, \eqref{particle_velocity_grains},  \eqref{diffusion_flux_tricrystal_3D}, and \eqref{ficks_law_3D} into  \eqref{local_inequality_gb1}, and neglecting the terms of the order of $|\dot{\boldsymbol{\mathcal C}}_a|^2$ and $|\dot{\boldsymbol R}_a\dot{\boldsymbol{\mathcal C}}_a|$, we can reduce the inequality to the form (no summation for repeated index $i$)
 \begin{equation}
V_i{f}_i+{\boldsymbol w}\cdot{\boldsymbol{\mathcal G}}_i+\dot{\boldsymbol{\mathcal C}}_i
\cdot{\boldsymbol{\mathcal H}}_i\geq 0 ~\text{on}~{\mathscr S}_i, 
\label{kinetic_eq2_gb3_tricrystal}
\end{equation}
where 
\begin{equation}
{\boldsymbol{\mathcal H}}_i= {\boldsymbol\sigma}{\boldsymbol n}_i, ~
{\boldsymbol{\mathcal G}}_a={\boldsymbol x}\times({\boldsymbol\sigma}{\boldsymbol n}_a
+\rho\mu{\boldsymbol n}_a)+{\boldsymbol A}_a^T{\boldsymbol D}_a^{-1}{\boldsymbol A}_a
{\boldsymbol w}+2{\boldsymbol\Upsilon}_a,~\text{and}~{\boldsymbol{\mathcal G}}_3 = {\boldsymbol 0}.
\label{kinetic_relation_coeffs1}
\end{equation}
In the above relations, ${\boldsymbol D}_i^{-1}$ is the Moore-Penrose pseudoinverse of ${\boldsymbol D}_i$ 
which satisfies ${\boldsymbol D}_i^{-1}{\boldsymbol D}_i={\boldsymbol D}_i{\boldsymbol D}_i^{-1}
={\boldsymbol P}_i$ \cite{gupta1},
and ${\boldsymbol\Upsilon}_a$ is the axial vector of skew part of $(\partial_{{\boldsymbol\Theta}_a}
 \gamma_a){\boldsymbol\Theta}_a^T$. 
Considering the Onsager's reciprocity theorem \cite{onsager1}, we use \eqref{kinetic_eq2_gb3_tricrystal}
to postulate linear kinetic relations associated with ${\mathscr S}_a$ ($a=1,2$). In writing them we assume the relative translational velocities $\dot{\boldsymbol{\mathcal C}}_a$ to be decoupled from GB migration and the rotation rate of grain $G_1$; a theory without this assumption can easily be constructed along similar lines. The kinetic relations are taken as
\begin{equation}
V_a={\mathcal M}_1^{(a)}{f}_a+{\boldsymbol{\mathcal M}}_2^{(a)}
\cdot{\boldsymbol{\mathcal G}}_a,
\label{kinetic_3D1a}
\end{equation}
\begin{equation}
{\boldsymbol w}={\boldsymbol{\mathcal M}}_2^{(a)}{f}_a+{\boldsymbol{\mathcal M}}_3^{(a)}
{\boldsymbol{\mathcal G}}_a,
\label{kinetic_3D2a}
\end{equation}
\begin{equation}
\dot{\boldsymbol{\mathcal C}}_a = {\boldsymbol{\mathcal L}}_a{\boldsymbol{\mathcal H}}_a,
\label{kinetic_3D3a}
\end{equation}
where ${\mathcal M}_1^{(a)}$, ${\boldsymbol
{\mathcal M}}_2^{(a)}$, ${\boldsymbol{\mathcal M}}_3^{(a)}$, and ${\boldsymbol{\mathcal L}}_a$ are various kinetic coefficients, to be discussed next. Assuming ${\mathcal M}_1^{(a)}$ to be non-vanishing we eliminate ${f}_a$ from 
\eqref{kinetic_3D2a} using \eqref{kinetic_3D1a} to rewrite ${\boldsymbol w}$ as
\begin{equation}
{\boldsymbol w}=\tilde{\boldsymbol\beta}_a V_a+\tilde{\boldsymbol{\mathcal S}}_a{\boldsymbol{\mathcal G}}_a,
\label{kinetic_3D2a2}
\end{equation}
where ${\mathcal M}_a={\mathcal M}_1^{(a)}$, $\tilde{\boldsymbol\beta}_a={\boldsymbol
{\mathcal M}}_2^{(a)}/{\mathcal M}_a$, and $\tilde{\boldsymbol{\mathcal S}}_a={\boldsymbol
{\mathcal M}}_3^{(a)}-{\mathcal M}_a\tilde{\boldsymbol\beta}_a\otimes\tilde{\boldsymbol\beta}_a$. 
The coefficients ${\mathcal M}_a$, $\tilde{\boldsymbol\beta}_a$, and $\tilde{\boldsymbol{\mathcal S}}_a$
have the same physical interpretation as described in Section \ref{examples_junc}. Substituting \eqref{kinetic_3D1a}, 
\eqref{kinetic_3D3a}, and \eqref{kinetic_3D2a2} back into the dissipation 
inequality \eqref{kinetic_eq2_gb3_tricrystal} we derive the following restrictions:
${\mathcal M}_a>0$; $\tilde{\boldsymbol{\mathcal S}}_a$ and ${\boldsymbol{\mathcal L}}_a$ 
are symmetric positive semi-definite. In terms of these new kinetic coefficients \eqref{kinetic_3D1a} takes the form
\begin{equation}
V_a={\mathcal M}_a({f}_a+\tilde{\boldsymbol\beta}_a\cdot{\boldsymbol{\mathcal G}}_a).
\label{kinetic_3D1a1}
\end{equation}
Furthermore, when sliding is active and $\tilde{\boldsymbol{\mathcal S}}_a$
is invertible, we can eliminate ${\boldsymbol{\mathcal G}}_a$ from \eqref{kinetic_3D1a1} with 
the help of \eqref{kinetic_3D2a2} to obtain
\begin{equation}
V_a=\frac{{\mathcal M}_a}{1+{\mathcal M}_a\tilde{\boldsymbol\beta}_a
\cdot {\tilde{\boldsymbol{\mathcal S}}_a}^{-1}\tilde{\boldsymbol\beta}_a}\left({f}_a
+\tilde{\boldsymbol\beta}_a\cdot{\tilde{\boldsymbol{\mathcal S}}_a}^{-1}{\boldsymbol w}\right).
\label{kinetic_normal_velo_3D}
\end{equation}
 Using \eqref{kinetic_relation_coeffs1} and \eqref{kinetic_3D1a1}
 in \eqref{kinetic_3D2a2}, and rearranging the  resulting expression, we can derive
\begin{equation}
\left({\boldsymbol I}-\tilde{\boldsymbol{\mathcal Z}}_a{\boldsymbol A}_a^T{\boldsymbol D}_a^{-1}
{\boldsymbol A}_a\right){\boldsymbol w}={\mathcal M}_a{f}_a\tilde{\boldsymbol\beta}_a
+\tilde{\boldsymbol{\mathcal Z}}_a({\boldsymbol x}_a\times{\boldsymbol\sigma}{\boldsymbol n}_a
+2{\boldsymbol\Upsilon}_a+\rho\mu{\boldsymbol x}_a\times{\boldsymbol n}_a),
\label{kinetic_equations_interm1}
\end{equation}
where $\tilde{\boldsymbol{\mathcal Z}}_a={\tilde{\boldsymbol{\mathcal S}}_a}+{\mathcal M}_a
\tilde{\boldsymbol\beta}_a\otimes \tilde{\boldsymbol\beta}_a$. 
For an invertible $\tilde{\boldsymbol{\mathcal S}}_a$ we multiply both sides of
\eqref{kinetic_equations_interm1} by $\tilde{\boldsymbol{\mathcal Z}}_a^{-1}$ and 
integrate the result over ${\mathscr S}_1$ and ${\mathscr S}_2$ for $a=1$ and $a=2$, respectively. We add the two expressions to obtain the following expression for ${\boldsymbol w}$:
\begin{equation}
{\boldsymbol w}=\left(\sum_{a=1}^2\int_{{\mathscr S}_a}(
\tilde{\boldsymbol{\mathcal Z}}_a^{-1}-{\boldsymbol A}_a^T{\boldsymbol D}_a^{-1}
{\boldsymbol A}_a)\,da\right)^{-1} \sum_{a=1}^2\int_{{\mathscr S}_a}\left({\mathcal M}_a{f}_a
\tilde{\boldsymbol{\mathcal Z}}_a^{-1}\tilde{\boldsymbol\beta}_a
+{\boldsymbol x}_a\times{\boldsymbol\sigma}{\boldsymbol n}_a+2{\boldsymbol\Upsilon}_a\right)da,
\label{kinetic_equations_interm2}
\end{equation}
where we have used $\sum_{a=1}^2\int_{{\mathscr S}_a}\rho\mu\,{\boldsymbol x}_a\times{\boldsymbol n}_a
\,da=\rho \int_{G_1}{\boldsymbol x}\times
\nabla{\mu}\,dv={\boldsymbol 0}$ (the first equality follows from the 
divergence theorem, whereas the second equality holds due to vanishing diffusional flux in the grain). To summarize the results obtained so far, we have the governing equations for the motion of ${\mathscr S}_1$ and ${\mathscr S}_2$ in \eqref{kinetic_normal_velo_3D},
and the governing equation for rotation of $G_1$ in \eqref{kinetic_equations_interm2}.

The kinetic relations for ${\mathscr S}_3$ can be derived similarly. In doing so, however, we allow for translational velocity to couple with the normal motion. Starting with \eqref{kinetic_eq2_gb3_tricrystal}, and assuming linear kinetics, we obtain  
\begin{equation}
\dot{\boldsymbol{\mathcal C}}_3={\boldsymbol\beta}_3V_3+{\boldsymbol{\mathcal L}}_3
{\boldsymbol{\mathcal H}}_3,
\label{kinetic_3D24}
\end{equation}
\begin{equation}
V_3={\mathcal M}_3({f}_3+{\boldsymbol\beta}_3\cdot{\boldsymbol{\mathcal H}}_3),
\label{kinetic_3D14} 
\end{equation}
where ${\mathcal M}_3>0$, ${\boldsymbol\beta}_3$, and ${\boldsymbol{\mathcal L}}_3$ 
(positive semi-definite) are the mobility, geometric coupling factor, and sliding coefficient, 
respectively, for ${\mathscr S}_3$. Replacing $V_3$ from \eqref{kinetic_3D14} in \eqref{kinetic_3D24}
the expression for translational velocity can be rewritten as
\begin{equation}
\dot{\boldsymbol {\mathcal C}}_3={\mathcal M}_3{f}_3{\boldsymbol\beta}_3+
{\boldsymbol{\mathcal Z}}_3{\boldsymbol{\mathcal H}}_3,
\label{kinetic_3D25}
\end{equation}
where ${\boldsymbol{\mathcal Z}}_3={\boldsymbol{\mathcal L}}_3+{\mathcal M}_3{\boldsymbol\beta}_3
\otimes{\boldsymbol\beta}_3$. 
Since the translational velocities are homogeneous, and the outer grains have been assumed to be much larger than the embedded grain, we integrate \eqref{kinetic_3D25} over ${\mathscr S}_3$ to write 
\begin{equation}
\dot{\boldsymbol{\mathcal C}}_3=\int_{{\mathscr S}_3}\left({\mathcal M}_3{f}_3
{\boldsymbol\beta}_3+{\boldsymbol{\mathcal Z}}_3{\boldsymbol{\mathcal H}}_3\right)da.
\label{kinetic_3D26}
\end{equation}
On the other hand, the governing equation for average translation velocity of the embedded grain can be obtained by first integrating \eqref{kinetic_3D3a} for $a=1$ and $2$, respectively, and then adding them to obtain 
\begin{equation}
\dot{\boldsymbol C}_1=\frac{1}{\text{area}({\mathscr S}_1\cup{\mathscr S}_2)}\left(
\text{area}({\mathscr S}_1)\dot{\boldsymbol{\mathcal C}}_3 +
\sum_{a=1}^2\int_{{\mathscr S}_a}{\boldsymbol{\mathcal L}}_a{\boldsymbol{\mathcal H}}_a da\right),
\label{kinetic_3D27}
\end{equation}
where $\dot{\boldsymbol{\mathcal C}}_3$ is given by \eqref{kinetic_3D26}. In \eqref{kinetic_3D14} we have the governing equation for the normal motion of ${\mathscr S}_3$ and in \eqref{kinetic_3D27} for the translation of the embedded grain.

Finally, we derive the kinetic relations which govern junction dynamics. For negligible diffusion along the junction curve, the dissipation inequality \eqref{local_inequality_junc2} simplifies to ${\boldsymbol{\mathcal F}}_J\cdot{\boldsymbol q}_p\geq 0$. Assuming linear kinetics we 
postulate that 
\begin{equation}
{\boldsymbol q}_p= {\boldsymbol{\mathcal M}}_J{\boldsymbol{\mathcal F}}_J,
\label{junction_kinetics_3D}
\end{equation}
where ${\boldsymbol{\mathcal M}}_J$ is the positive semi-definite junction mobility tensor. An analogous treatment in a 2D setting can be seen in \cite{fischer1} (see also \cite{basak2}). 
The junction force ${\boldsymbol{\mathcal F}}_J$ given by 
\eqref{junction_force_expression_3D2} is a function 
of the unknown local orientations of the adjacent GBs which, for a non-splitting 
junction, can be calculated using the compatibility conditions $V_i={\boldsymbol q}_p\cdot{\boldsymbol n}_i$ 
(see \cite{fischer1,basak2} for a detailed calculation in 2D).

To conclude, the complete set of kinetic equations governing the coupled GB motion 
in the tricrystalline arrangement includes \eqref{kinetic_normal_velo_3D} for the 
motion of ${\mathscr S}_1$ and ${\mathscr S}_2$, \eqref{kinetic_3D14} for the motion of  
${\mathscr S}_3$, \eqref{kinetic_equations_interm2} for the rotation of the embedded grain $G_1$ (the outer grains are non-rotating), \eqref{kinetic_3D27} and 
\eqref{kinetic_3D26} for the translation of grains $G_1$ and $G_2$, respectively (whereas grain $G_3$ is stationary), and 
\eqref{junction_kinetics_3D} (in association with the compatibility condition) for the motion of the junction curve.

\section{Concluding remarks}
\label{conclusion_junctions_3D}
We have presented a thermodynamically consistent 3D study of coupled GB motion in the 
presence of junctions, hitherto restricted to 2D crystalline materials. Towards this end we introduced a novel 
continuum mechanics based theory of irreversible dynamics of incoherent interfaces with junctions, which allows for diffusion in the bulk, on the interface, and along the junction curve. The various local dissipation inequalities derived therein were used to motivate kinetic relations for the coupled GB motion in two
bicrystals and one tricrystal. These relations were solved analytically whenever it was possible to do so, but were otherwise left in a form amenable to numerical computations. In any case, the results clearly demonstrated the effect of coupling on the grain dynamics. Consider for instance the shrinking of an isolated grain, embedded within a larger grain, under the action of capillary. Without coupling, the embedded grain can disappear only by shrinking to a vanishing size. However with coupling, the grain can disappear by aligning its orientation with the outer grain even before it has shrunk significantly \cite{basak1}. The proposed kinetic relations also emphasize the coupling of junction dynamics with both grain and GB motion. Depending on the junction mobility, grain dynamics can experience a substantial drag compared to the case with no junctions \cite{basak2}.

The present work can form a basis for research in several future directions. The theory of incoherent interfaces, which includes junctions and diffusion, is in fact applicable to a more general situation where the grains are allowed to deform plastically. The resulting framework would be useful for phenomena which involves coupling of plastically deforming bulk with moving incoherent interfaces and junctions. Secondly, the kinetic relations derived for the tricrystal can be used to study the coupled motion in  
polycrystalline materials containing large number of grains and junctions. Of course, this will demand significant computational effort and hence efficient numerical algorithms. A related direction of work would be to develop numerical techniques (such as level set methods) for solving equations of coupled motion of an embedded grain with anisotropic constitutive properties. Thirdly,  the computation of the vectorial coupling
factor, which has been restricted here to small angle GBs, should be extended
to large angle GBs. Finally, 3D atomistic simulations will be required to clarify the nature of various kinetic coefficients (including the coupling factor), in particular regarding their dependence on three misorientation angles and two orientation angles of the GB. 

\bibliography{junction_3D}
\bibliographystyle{plain}   

\end{document}